\documentclass[10pt,letterpaper]{article}
\usepackage[margin=2cm]{geometry}

\usepackage{amsmath,amssymb}

\usepackage{changepage}

\usepackage[utf8x]{inputenc}

\usepackage{textcomp,marvosym}

\usepackage{cite}

\usepackage{nameref,hyperref}

\usepackage{microtype}
\DisableLigatures[f]{encoding = *, family = * }

\usepackage[table]{xcolor}

\usepackage{array}

\newcolumntype{+}{!{\vrule width 2pt}}

\newlength\savedwidth

\usepackage[aboveskip=1pt,labelfont=bf,labelsep=period,justification=raggedright,singlelinecheck=off]{caption}

\makeatletter
\renewcommand{\@biblabel}[1]{\quad#1.}
\makeatother

\usepackage{lastpage,fancyhdr,graphicx}
\usepackage{epstopdf}
\fancyheadoffset[L]{2.25in}
\fancyfootoffset[L]{2.25in}
\newcommand{\beginsupplement}{%
        \setcounter{table}{0}
        \renewcommand{\thetable}{S\arabic{table}}%
        \setcounter{figure}{0}
        \renewcommand{\thefigure}{S\arabic{figure}}%
        \setcounter{equation}{0}
        \renewcommand{\theequation}{S\arabic{equation}}
     }

\usepackage{xr}

  \graphicspath{{./Fig_f/},{For_revision/}}

\usepackage{epstopdf}

\usepackage{subfig}

\usepackage{enumitem}

\begin{document}
\vspace*{0.2in}

\begin{flushleft}
{\Large
\textbf\newline{Adapt or perish: Evolutionary rescue in a gradually deteriorating environment} 
}
\newline
\\
Loïc Marrec\textsuperscript{1},
Anne-Florence Bitbol\textsuperscript{1,2*}
\\
\bigskip
\textbf{1} Sorbonne Universit{\'e}, CNRS, Institut de Biologie Paris-Seine, Laboratoire Jean Perrin (UMR 8237), F-75005 Paris, France\\
\textbf{2} Institute of Bioengineering, School of Life Sciences, École Polytechnique Fédérale de Lausanne (EPFL), CH-1015 Lausanne, Switzerland
\bigskip

%
%


* anne-florence.bitbol@epfl.ch

\end{flushleft}
\section*{Abstract}

We investigate the evolutionary rescue of a microbial population in a gradually deteriorating environment, through a combination of analytical calculations and stochastic simulations. We consider a population destined for extinction in the absence of mutants, which can only survive if mutants sufficiently adapted to the new environment arise and fix. We show that mutants that appear later during the environment deterioration have a higher probability to fix. The rescue probability of the population increases with a sigmoidal shape when the product of the carrying capacity and of the mutation probability increases. Furthermore, we find that rescue becomes more likely for smaller population sizes and/or mutation probabilities if the environment degradation is slower, which illustrates the key impact of the rapidity of environment degradation on the fate of a population. We also show that our main conclusions are robust across various types of adaptive mutants, including specialist and generalist ones, as well as mutants modeling antimicrobial resistance evolution. We further express the average time of appearance of the mutants that do rescue the population and the average extinction time of those that do not. Our methods can be applied to other situations with continuously variable fitnesses and population sizes, and our analytical predictions are valid in the weak-to-moderate mutation regime.



\section*{Introduction}

Understanding how a population of living organisms can survive in a gradually deteriorating environment is a fundamental question in evolution~\cite{Waxman11,Uecker11, Peischl12}, which is particularly relevant in the pressing context of climate change~\cite{Bell09,Chevin10,Pauls13,Botero15,Nadeau17}. Addressing this question is also important in order to understand antimicrobial resistance evolution, which often occurs in a variable environment, as antimicrobial is added to a medium or given to a patient~\cite{Lin16,Levin-Reisman17}. In fact, even when antimicrobial is added instantaneously, yielding an abrupt environment switch, the resulting fitness decrease is gradual~\cite{Lin16}. In a deteriorating environment, the fitness of wild-type organisms decreases with time. In the simple case of asexual microorganisms, their division rate can then become smaller than their death rate, which yields a decrease of population size, eventually leading to extinction~\cite{Coates18}. However, the population can be rescued by a mutation which is better adapted to the new environment, and restores positive population growth (or several such mutations): this phenomenon is called evolutionary rescue~\cite{Martin12,Gonzalez13,Alexander14,Carlson14,Barton18}.

A gradually deteriorating environment impacts the population size and the fitness of the wild-type organism, which can both strongly impact the fate of a mutation~\cite{Uecker11}. The decay of the wild-type population simultaneously entails a decreased frequency of mutant appearance, which can hinder rescue, and a decreased competition for existing mutants, known as competitive release~\cite{Wargo07,Kouyos14}, which can facilitate rescue. Studying the evolutionary rescue of a population in a gradually deteriorating environment requires  accounting for simultaneous continuous time variations of fitness, population size and population composition, which makes it complex. Varying patterns of selection have recently been the focus of significant interest, mainly in the case of switches between different environment states, highlighting their strong effect on evolution~\cite{Kussell06,Mustonen08,Rivoire11,Melbinger15,Cvijovic15,Skanata16,Hufton16,Wienand17,Mayer17,Meyer18,Danino18,Marrec18,Trubenova19,Marrec20}. Despite its practical relevance, the case of a fitness varying continuously over time for a given genotype has been comparatively less studied, with a focus on stabilizing selection~\cite{Burger95,Gomulkiewicz09,Kopp07,Kopp09a,Kopp09b,Matuszewski14,Alexander14} or on the fate of a single beneficial mutation~\cite{Waxman11, Uecker11, Peischl12}. Furthermore, most theoretical works on evolutionary rescue consider an abrupt environment change~\cite{Orr08,Bell09,Martin12,Anciaux18}. Here we address evolutionary rescue in a gradually changing environment, which deteriorates from the point of view of wild-type organisms.

Adaptation to a new environment can occur in multiple ways. A specialist mutant that is particularly well-adapted to this new environment can emerge, e.g. a thermophilic mutant in the case of a temperature rise. Another possibility is the appearance of a generalist mutant, which is able to grow in both the initial and the final environments, while being less fit than specialists in their respective favorite environments~\cite{Donaldson08,Mayer17,Wang19,SachdevaPreprint}. Yet another one regards mutants that are less fit in the final environment than in the initial one, but still sufficiently fit to be able to grow in the final environment. The latter case can model the evolution of antimicrobial resistance as drug concentration is increased from zero to a value that is above the minimum inhibitory concentration of the sensitive microbes but below that of the resistant microbes~\cite{Gullberg11,Yu18}.

In the present work, we consider a microbial population subjected to a gradual environment deterioration, such that the fitness and the size of the wild-type population are gradually decaying, and that extinction would be certain in the absence of adaptation. We study the fixation probability of generalist and specialist adaptive mutants as a function of the time when they appear during the environment deterioration, and we also consider a model of antimicrobial resistance evolution. We obtain an expression for the overall probability that the population is rescued by an adaptive mutation, thereby avoiding extinction. We investigate the dependence of the rescue probability on the rapidity of the environment deterioration, as well as on population size and mutation probability. We also compare different types of mutants. We further express the average time of appearance of the mutants that do rescue the population and the average extinction time of those that do not.

\section*{Model and methods}  

\subsection*{Population model}

We consider a population of asexual microorganisms with carrying capacity $K$, corresponding to the maximum population size that the environment can sustain, given e.g. the nutrients available. We assume that two types of microorganisms can exist in this population: wild-type (W) and mutant (M). The division rate of each organism is assumed to be logistic \cite{Verhulst}, and reads $f_i(t)(1-N/K)$, where $N$ represents the total population size, while the time-dependent fitness $f_i(t)$ with $i=W$ or $i=M$ represents the maximal possible division rate of the (wild-type or mutant) organism at time $t$, which would be reached if $N\ll K$. The death rates of W and M organisms are respectively denoted by $g_W$ and $g_M$. Note that (Malthusian) fitness is usually measured as the exponential growth rate at the population scale, and that fitness often refers to the overall outcome of both survival and selection: under such definitions, fitness would in fact correspond to $f_i(t)-g_i$ here. However, here we will not consider any variability of death rates, and thus, for the sake of simplicity, fitness will refer to $f_i(t)$ throughout. While we assume that the variability of the environment impacts fitnesses and not death rates, our approach can be easily extended to variable death rates. We further assume that W microorganisms can mutate into M microorganisms with the mutation probability $\mu$ upon each division. We do not consider back mutations. Note that because mutations occur upon division, the number of mutants appearing per unit time depends both on the population size and on the fitness of W microorganisms. Importantly, our model incorporates both variations of population size (population dynamics) and of composition (population genetics)~\cite{Melbinger10,Melbinger15,Huang15}. Throughout, we consider the fitness of W microorganisms in the initial environment as the reference fitness and set it to 1. Therefore, our time unit corresponds to the inverse of this fitness (which is the maximum division rate we consider).

We start from a microbial population composed of $N_W(0)=N_{W}^0$ wild-type microorganisms and no mutant. Specifically, our simulations include a phase of initial growth, which can model e.g. the development of an infection starting from the bottleneck at transmission~\cite{Abel15}. In practice we will start our simulations with $N_{W}^0=10$. Fig.~\ref{Robust} demonstrates that our results do not depend on this particular choice, since starting with $N_{W}^0=10$ gives the same results as starting with $N_{W}^0=K[1-g_W/f_W(0)]$, which corresponds to the stationary population size in the initial environment within a deterministic description. Note however that if we started with a very small number of W microorganisms (i.e. 1 or 2), we would need to take into account rapid stochastic extinctions of the population~\cite{ovaskainen10}: we will not consider this regime.

\subsection*{Fitnesses in a deteriorating environment}

\begin{figure}[h!]
	\centering
\includegraphics[width=\textwidth]{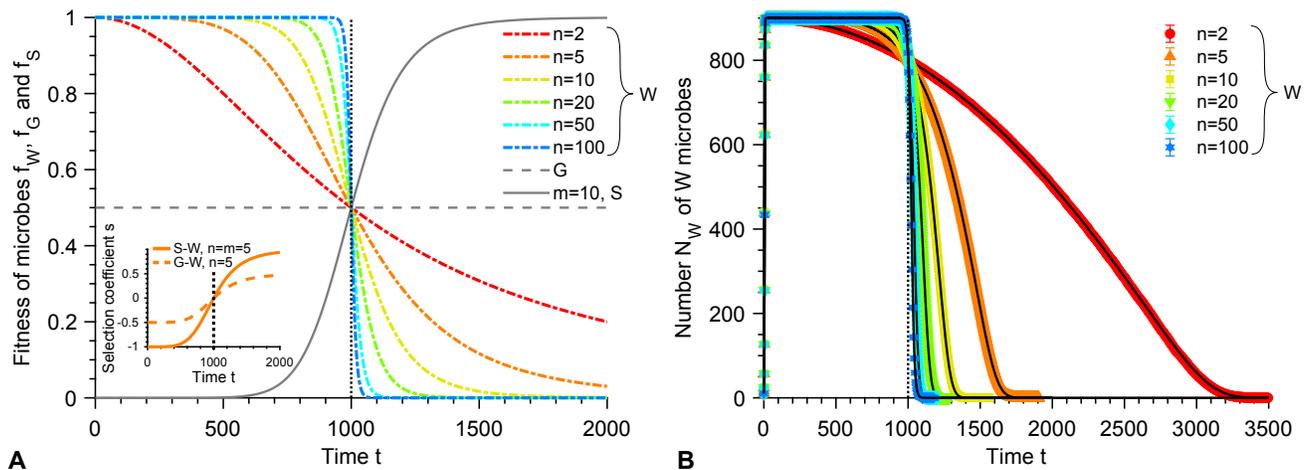}
	\vspace{0.2cm}
	\caption{{\bf Fitnesses and wild-type population in a deteriorating environment.} {\bf A:} Fitnesses $f_W$, $f_G$ and $f_S$ of the wild-type organisms (W), generalist (G) and specialist (S) mutants versus time $t$ (see Eqs.~\ref{HE_eq} and~\ref{fS}). Several values of the Hill coefficient $n$ are shown for W. Inset: selection coefficient $s=f_M-f_W$ for both types of mutants $M=G$ or $S$ versus time $t$, shown with $n=5$ (and $m=5$ for S mutants). {\bf B:} Number $N_W$ of W microbes versus time $t$ for different values of $n$ (same colors as in  {\bf A}). Data points correspond to averages over $10^3$ replicate stochastic simulations, and error bars (smaller than markers) represent 95\% confidence intervals. Black solid curves correspond to numerical integrations of Eq. \ref{dNAdt}. Parameter values: $g_W=g_S=g_G=0.1$, $K=10^3$, $N_W^0=10$, and $\theta=10^3$. Vertical dotted line in both panels: $t=\theta$.}%
	\label{HE_NA}%
\end{figure}

 To model the impact of a continuously deteriorating environment on the fitness of W microorganisms, we choose the Hill function: 
\begin{equation}
f_W(t)=\frac{1}{1+(t/\theta)^n}\mbox{ },
\label{HE_eq}
\end{equation}
where $n$ is the Hill coefficient and $\theta$ the inflection point, such that $f_W(\theta)=0.5$. This sigmoidal function represents a transition between two different environments, by decreasing from the reference fitness value $f_W(0)=1$ toward 0 as $t$ increases, with a steepness that is tunable via $n$. Specifically, the decay is more abrupt manner for larger values of $n$ (see Fig. \ref{HE_NA}A). The Hill function is quite generic in biological contexts, e.g. it is a good model for cooperative reactions, and for the pharmacodynamics of antimicrobials \cite{Regoes04}. Moreover, Eq.~\ref{HE_eq} allows us to recover the case of an abrupt environment change as a limiting case when $n\rightarrow\infty$. Because it is $n$ that sets the timescale of the environmental change occurring around $\theta$, we will vary $n$ at a fixed (and large) value of $\theta$. Note that employing Eq.~\ref{HE_eq} implies environment changes with rates symmetric with respect to $\theta$. But crucially, the methods presented here do not depend on the exact function chosen and can be applied to other forms of environment degradation beyond Eq.~\ref{HE_eq}. 

We will mainly consider two types of adaptive mutants. First, generalist mutants, denoted by G, are not impacted by gradual changes of the environment and have a constant fitness $f_G$. We choose $f_G=0.5$ so that G mutants and W organisms have the same time-averaged fitness. Second, specialist mutants, denoted by S, have a fitness described by an increasing Hill function, so that they are better adapted to the final environment, in contrast to W organisms:
\begin{equation}
f_S(t)=\frac{(t/\theta)^m}{1+(t/\theta)^m}\,.
\label{fS}
\end{equation} 
We take the same point of inflection $\theta$ for W and S, as it marks the midst of the environmental transition. Conversely, we allow different Hill coefficients $n$ and $m$, reflecting a different sensitivity of W and S individuals to environmental change (see Fig. \ref{HE_NA}A).  Note that S mutants, G mutants and W organisms have the same time-averaged fitness over a time window that is symmetric around $\theta$, and that G mutants are in fact S mutants with $m=0$. The selection coefficient, defined as the fitness difference between mutant and wild-type (see Fig.~\ref{HE_NA}A, inset), switches from negative to positive at the inflection point, more steeply when $n$ and $m$ are large, and with a wider range for S mutants than for G mutants. 

In section~\ref{antibio} of the Supporting Information, we also consider another type of mutant in order to model antimicrobial resistance evolution. We focus on the case where drug concentration is increased from zero to a value that is above the minimum inhibitory concentration of the sensitive microbes but below that of the resistant microbes~\cite{Gullberg11,Yu18}. Then, resistant mutants are able to grow in the final environment and rescue the population.

\subsection*{Methods}

We present both analytical and numerical results. Our analytical results are obtained using methods from stochastic processes, especially from birth-death processes with time varying rates~\cite{NissenMeyer66,Bailey,Uecker11,Alexander12,Parzen}. Importantly, our predictions make quite minimal assumptions and hold in the weak-to-moderate mutation regime where $K\mu\lesssim 1$. Our simulations employ a Gillespie algorithm~\cite{Gillespie76,Gillespie77}, and incorporate all individual stochastic division, mutation and death events with the associated rates. In principle, the time variability of the division rates imposes a difficulty~\cite{Thanh15}, but the short duration of time intervals between individual events allows us to neglect rate variations between events (see Supporting Information, section~\ref{SimuAppendix} for details). Our model allows us to fully account for the stochasticity of mutation occurrence and establishment~\cite{Ewens79,Rouzine01,Fisher07,Patwa08,Weissman09}, as well as that of population extinction~\cite{Coates18,Teimouri19,AlexanderPreprint}. Matlab implementations of our numerical simulations are freely available at \url{https://doi.org/10.5281/zenodo.3993272}.

In our analytical calculations, we will often make a deterministic approximation for the evolution of the number $N_W$ of W individuals, while the evolution of the mutant population will be described in a fully stochastic manner. Indeed, mutants are in small numbers when they appear, while they generally arise in a large population of W organisms. In the deterministic limit, $N_W$ satisfies the following ordinary differential equation:
\begin{equation}
\frac{\mbox{d}N_W}{\mbox{d}t}=\left[f_W(t)\left(1-\frac{N_W}{K}\right)-g_W\right]N_W\mbox{ }.
\label{dNAdt}
\end{equation}
This description is appropriate for very large $N_W$, and Eq.~\ref{dNAdt} can be derived from the complete stochastic model in this limit (see Supporting Information, Section \ref{From the stochastic model to the deterministic limit} and Refs.~\cite{VanKampen, gardiner}). 

Fig. \ref{HE_NA}B compares the predictions from Eqs. \ref{HE_eq} and \ref{dNAdt} to the results of stochastic simulations (see Supporting Information Section \ref{NS - Hill equation}), and demonstrates the validity of the deterministic approximation in this regime. Fig. \ref{HE_NA}b also illustrates that in the absence of mutants, the population of W individuals always goes extinct, due to the fact that fitness $f_W$ tends to 0 while death rate is nonzero ($g_W>0$). Moreover, the bigger the Hill coefficient $n$, the faster the W population goes extinct.

\section*{Results}

\subsection*{Fixation probability of mutants: on the importance of good timing}
\label{Fixation probability of mutants}

In a deteriorating environment, mutants will have different fates depending on when they appear. Therefore, before investigating overall rescue probabilities, we address the fixation probability $p_{\mbox{\scriptsize fix}}(t_0)$ of a mutant as a function of the time $t_0$ when it appears during the environment deterioration. Competition with wild-type organisms is felt by mutants through their division rate $f_M(t)\{1-[N_W(t)+N_M(t)]/K\}$.  
At the early stages when competition matters, i.e. when the logistic term is important, the number of mutants is small with respect to the number of wild-type microorganisms, $N_M(t)\ll N_W(t)$, and thus the division rate of mutants can be approximated by $f_M(t)[1-N_W(t)/K]$. Furthermore, at these early stages, the number of wild-type microorganisms $N_W$ is large enough to be described in a deterministic framework (see Models and Methods, Eq. \ref{dNAdt} and Fig.~\ref{HE_NA}). We retain a full stochastic description for mutants, which are in small numbers just after the mutation arises~\cite{Uecker11,Alexander12,Parzen}, and we introduce the probability $P(i, t|1, t_0)$ of having $i$ mutants at time $t$ knowing that there is 1 mutant at time $t_0$. The fixation probability of the mutants can then be obtained from the probability generating function $\phi(z,t)=\sum_{i=0}^{\infty}z^iP(i,t|1,t_0)$, which satisfies $p_{\mbox{\scriptsize fix}}(t_0)=1-\lim_{t\to\infty}P(0,t|1,t_0)=1-\lim_{t\to\infty}\phi(0,t)$. Solving the partial differential equation governing the evolution of $\phi(z,t)$ (see Supporting Information, section~\ref{probaFix}) yields~\cite{Uecker11,Alexander12,Parzen}
\begin{equation}
p_{\mbox{\scriptsize fix}}(t_0)=\frac{1}{1+g_M\int_{t_0}^{\infty} e^{\rho(t)}\mbox{d}t},
\label{pfix}
\end{equation}
where
\begin{equation}
\rho(t)=\int_{t_0}^t\left[g_M-f_M(u)\left(1-\frac{N_W(u)}{K}\right)\right]\mbox{d}u\mbox{ }.
\label{rho}
\end{equation}
Numerical integration of Eq. \ref{pfix} is discussed in section \ref{Numerical resolution details} of the Supporting Information.

\begin{figure}[h!]%
	\centering
	\includegraphics[width=\textwidth]{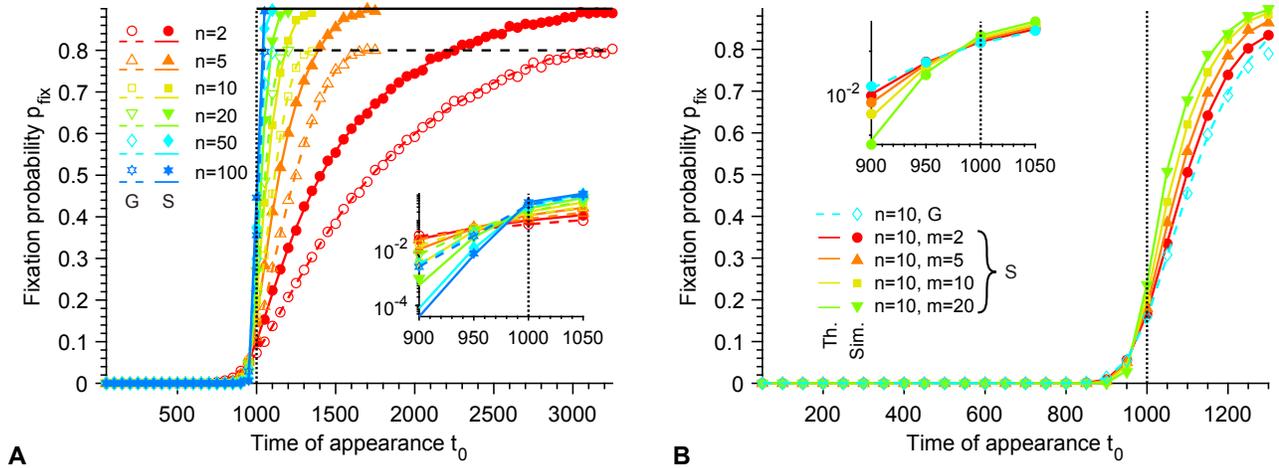}
	\vspace{0.2cm}
	\caption{{\bf Fixation probability of mutants.} {\bf A.} Fixation probability $p_{\mbox{\scriptsize fix}}$ of G and S mutants versus their time of appearance $t_0$ in the deteriorating environment, for different Hill coefficients $n$ characterizing the steepness of the environment deterioration (see Eq.~\ref{HE_eq}). Here, S mutants satisfy $m=n$, i.e. they have the same sensitivity to the environment as W organisms (see Eq.~\ref{fS}). Horizontal dashed line: $p_{\mbox{\scriptsize fix}}=1-g_G/f_G$. Horizontal solid line: $p_{\mbox{\scriptsize fix}}=1-g_S$. Data is shown for $t_0<\tau_W$, where $\tau_W$ is the average extinction time of the W population in the absence of mutation.  {\bf B.} Fixation probability $p_{\mbox{\scriptsize fix}}$ of different types of mutants versus their time of appearance $t_0$ in the deteriorating environment, for a fixed Hill coefficient $n=10$ characterizing the decay of $f_W$ (see Eq.~\ref{HE_eq}). G mutants and S mutants with different Hill coefficients $m$ (see Eq.~\ref{fS}), corresponding to different sensitivities to the changing environment, are considered. In both panels, markers correspond to averages over $10^4$ replicate stochastic simulations (``Sim.''). Dashed and solid lines correspond to numerical integrations of Eq. \ref{pfix} (``Th.'') for G and S mutants, respectively. Parameter values: $g_W=g_G=g_S=0.1$, $K=10^3$, $N_W^0=10$ and $\theta=10^3$. Vertical dotted lines: $t_0=\theta$. Main panels: linear scale; insets: semi-logarithmic scale. }%
	\label{HE_pfix}%
\end{figure}

Fig. \ref{HE_pfix} shows the fixation probability $p_{\mbox{\scriptsize fix}}$ of a mutant versus the time $t_0$ at which it appears during the deterioration of the environment. A very good agreement is obtained between the results of our stochastic simulations and the analytical prediction of Eq. \ref{pfix}. This holds both when $t_0<\theta$, while mutants are less fit than W organisms, and when $t_0>\theta$, where the opposite is true. In Fig.~\ref{MoreG}, we provide additional results for the fixation probability of generalist mutants with different fitness values $f_G$, which thus become effectively beneficial sooner or later during the environment deterioration, illustrating that Eq.~\ref{pfix} holds in these various cases. 

Fig. \ref{HE_pfix} shows that $p_{\mbox{\scriptsize fix}}$ strongly increases with $t_0$: mutants appearing later in the environmental degradation are much more likely to fix. This reflects both the increasing intrinsic fitness advantage of mutants due to the environment transition, and the decreasing competition with the W population that decays as the environment deteriorates for W organisms. Note that variations of selection coefficients only, or of competition pressure only, were previously addressed~\cite{Uecker11}, and that an increase in fixation probability with mutant appearance time was described under decreasing competition~\cite{Alexander12}. Fig. \ref{HE_pfix}A shows that the increase of $p_{\mbox{\scriptsize fix}}$ is strong around the inflection point $\theta$, and is steeper for larger Hill coefficients $n$ characterizing the fitness decay of the wild-type organisms (see Eq.~\ref{HE_eq}). Furthermore, for each value of $n$, sufficiently before $\theta$, generalist  (G) mutants are more likely to fix than specialist (S) mutants with $m=n$ (see Models and Methods, Eq.~\ref{fS}), because then $f_G>f_S$. Conversely, S mutants are more likely to fix than G mutants sufficiently after $\theta$ because $f_G<f_S$. Note that in section \ref{Approx} of the Supporting Information, we provide analytical approximations for the fixation probability with large Hill coefficients $n,m \rightarrow \infty$. Finally, Fig. \ref{HE_pfix}B shows that for $t_0>\theta$, $p_{\mbox{\scriptsize fix}}$ increases with the Hill coefficient $m$ characterizing the steepness of the fitness transition for S mutants, and all S mutants are more likely to fix than G mutants, consistently with the fact that G mutants correspond to S mutants with $m=0$ (see Eq.~\ref{fS}).

For large $t_0$, if the W population is not extinct yet, the fixation probability $p_{\mbox{\scriptsize fix}}$ in Eq.~\ref{pfix} converges to $1-g_G/f_G$  (resp. $1-g_S$) for G (resp. S) mutants, which is corroborated by our simulation results (see Figs.~\ref{HE_pfix}A and~\ref{MoreG}A). This simple limit can be interpreted as follows: mutants appearing just before the extinction of the W population face negligible competition, and thus they survive and fix unless they undergo rapid stochastic extinction~\cite{ovaskainen10, Coates18,Marrec20}. Note that $p_{\mbox{\scriptsize fix}}$ is constructed so that mutant lineages that undergo rapid stochastic extinctions are counted as not fixing in the population. Importantly, even though the fixation probability $p_{\mbox{\scriptsize fix}}$ at a given $t_0$ becomes larger as $n$ is increased, mutants appearing just before the extinction of the W population (which occurs faster as $n$ is increased, see Fig.~\ref{HE_NA}B) have a fixation probability independent of $n$ (see Figs.~\ref{HE_pfix}A and~\ref{pfix_rescaled}).

\newpage

\subsection*{Rescue probability}
\label{Probability of appearance and fixation of mutants}

\begin{figure}[h!]%
	\centering
	\includegraphics[width=\textwidth]{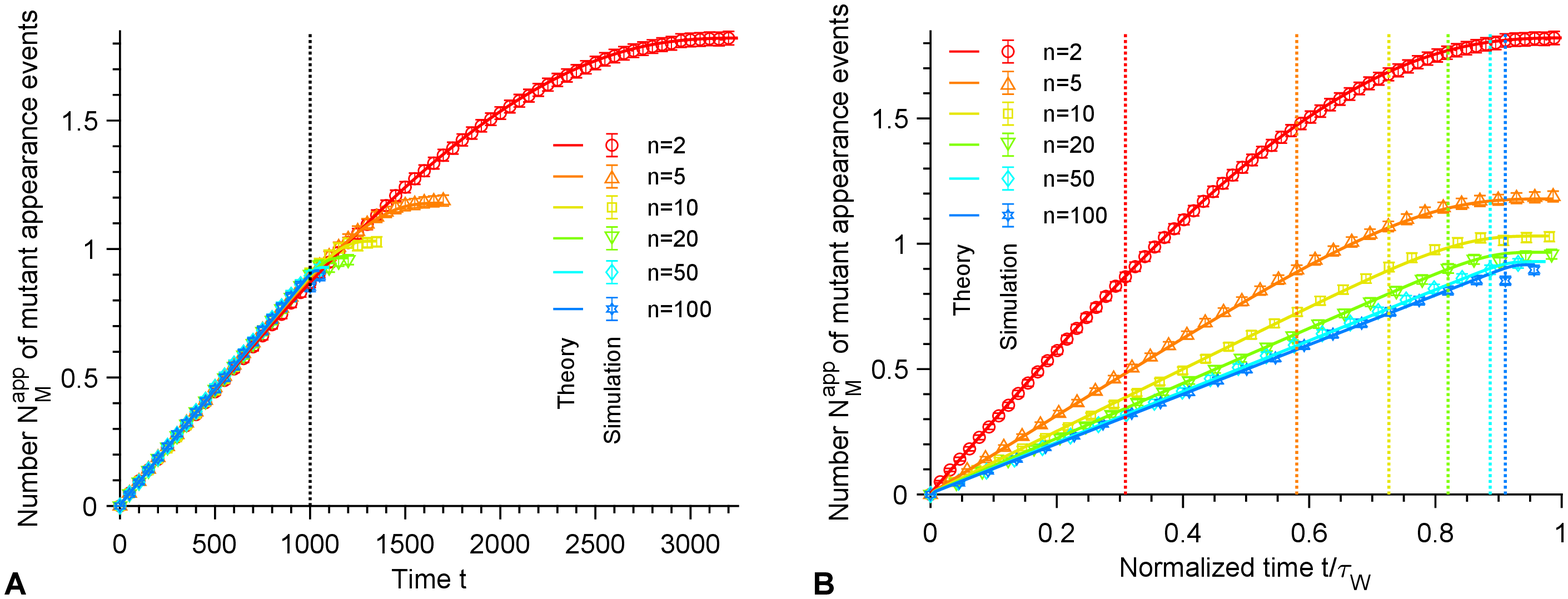}%
	\vspace{0.2cm}
	\caption{{\bf Appearance of mutants.} {\bf A.} Average number $N_M^{app}$ of mutant appearance events that can occur between times 0 and $t$, plotted versus time $t$, for different Hill coefficients $n$ characterizing the steepness of the environment deterioration. Vertical dotted line: $t=\theta$. Markers correspond to averages over $10^4$ replicate stochastic simulations (``Simulation''), where mutants that appear are replaced immediately by wild-type organisms to avoid any mutant fixation events and count all potential mutant appearance events. Solid lines correspond to numerical integrations of $N_M^{app}(t)=\int_0^\infty N_W(t)f_W(t)(1-N_W(t)/K)\mu \mbox{d}t$ (``Theory''), which corresponds to the number of mutants that appear, assuming that $N_M(t)\ll N_W(t)$ when they appear (see main text above Eq.~\ref{dpaf1}). {\bf B.} Same data, rescaled by the average extinction time $\tau_W$ of the wild-type population in the absence of mutation. Vertical dotted lines: $t=\theta$. Parameter values: $g_W=0.1$, $K=10^3$, $\theta=10^3$, $\mu=10^{-5}$ and $N_W^0=10$. Data is shown for $t<\tau_W$.
	}%
	\label{NMFig}%
\end{figure}

So far, we investigated the fate of a given mutant lineage as a function of its appearance time during the environment degradation. Let us now address whether mutants can rescue the population or not. For a mutation probability $\mu$ at division, both the occurrence of a new mutation and its subsequent fixation probability depend on the number and division rate of W organisms. We thus consider the probability $p_{\mbox{\scriptsize af}}(t)$ that a mutant appears between 0 and $t$ and fixes. The rescue probability $p_{\mbox{\scriptsize r}}$ corresponds to the probability that a mutant appears and fixes before the microbial population goes extinct, and is thus given by $p_{\mbox{\scriptsize r}}=\underset{t \rightarrow \infty}{\lim}p_{\mbox{\scriptsize af}}(t)$. Using Bayes' rule, the probability that a mutant appears between $t$ and $t+\mbox{d}t$ and fixes (which is equal to the probability that no mutant destined for fixation appeared before, and that a mutant destined for fixation then appears), denoted by $\mbox{d}p_{\mbox{\scriptsize af}}(t)=p_{\mbox{\scriptsize af}}(t+\mbox{d}t)-p_{\mbox{\scriptsize af}}(t)$, can be written as:
\begin{equation}
\mbox{d}p_{\mbox{\scriptsize af}}(t)=(1-p_{\mbox{\scriptsize af}}(t))\mbox{d}p_{\mbox{\scriptsize naf}}(t) \mbox{ },
\end{equation} 
where $(1-p_{\mbox{\scriptsize af}}(t))$ is the probability that no mutant destined for fixation appeared before, while $\mbox{d}p_{\mbox{\scriptsize naf}}(t)$ is the probability that a mutant appears between $t$ and $t+\mbox{d}t$ and fixes, provided that no mutant destined for fixation appeared before. The latter can be calculated by considering that the population is fully or mostly wild-type at time $t$, i.e. $N_W(t)\gg N_M(t)$, which is expected to be valid in most cases, except in the strong-mutation regime $K\mu\gg 1$ where multiple mutant lineages arise almost simultaneously. Then, $\mbox{d}p_{\mbox{\scriptsize naf}}(t)=p_{\mbox{\scriptsize fix}}(t)\mbox{d}N_M^{app}(t)$, where $\mbox{d}N_M^{app}(t)=N_W(t)f_W(t)(1-N_W(t)/K)\mu \mbox{d}t$ is the number of mutants that appear between $t$ and $t+\mbox{d}t$ in a fully wild-type population (see Fig.~\ref{NMFig}). Thus,
\begin{equation}
\frac{\mbox{d}p_{\mbox{\scriptsize af}}(t)}{1-p_{\mbox{\scriptsize af}}(t)}=p_{\mbox{\scriptsize fix}}(t)N_W(t)f_W(t)\left(1-\frac{N_W(t)}{K}\right)\mu \mbox{d}t \mbox{ }.
\label{dpaf1}
\end{equation} 
We again take a deterministic description for $N_W(t)$ (see Eq.~\ref{dNAdt}), and the fitness $f_W(t)$ of W organisms is given by Eq.~\ref{HE_eq}. Then, integrating Eq. \ref{dpaf1} with $p_{\mbox{\scriptsize af}}(0)=0$ yields $p_{\mbox{\scriptsize af}}(t)=1-\exp\left[-\sigma(t)\right]$, with
\begin{equation}
\sigma(t)=\mu\int_0^{t}p_{\mbox{\scriptsize fix}}(s)N_W(s)f_W(s)\left(1-\frac{N_W(s)}{K}\right) \mbox{d}s \mbox{ }.
\label{petitsigma}
\end{equation}
Taking the limit $t\rightarrow\infty$ then gives the rescue probability
\begin{equation}
p_{\mbox{\scriptsize r}}=\underset{t \rightarrow \infty}{\lim}p_{\mbox{\scriptsize af}}(t)=1-\exp\left(-\Sigma\right) \mbox{ },
\label{paf}
\end{equation}
where
\begin{equation}
\Sigma=\underset{t \rightarrow \infty}{\lim}\sigma(t)=\mu\int_0^{\infty}p_{\mbox{\scriptsize fix}}(t)N_W(t)f_W(t)\left(1-\frac{N_W(t)}{K}\right) \mbox{d}t \mbox{ }.
\label{theSigma}
\end{equation}
Note that if $\Sigma \ll 1$, Eq.~\ref{paf} reduces to $p_{\mbox{\scriptsize r}} \approx \Sigma$, which would be obtained by neglecting possible earlier fixations. Note also that, since mutant lineages undergoing rapid stochastic extinction are counted as not fixing in $p_{\mbox{\scriptsize fix}}$, they are correctly counted as not able to rescue the population. Numerical integration of Eqs. \ref{paf}-\ref{theSigma} is discussed in section \ref{Numerical resolution details} of the Supporting Information.

\begin{figure}[h!]%
	\centering
	\includegraphics[width=\textwidth]{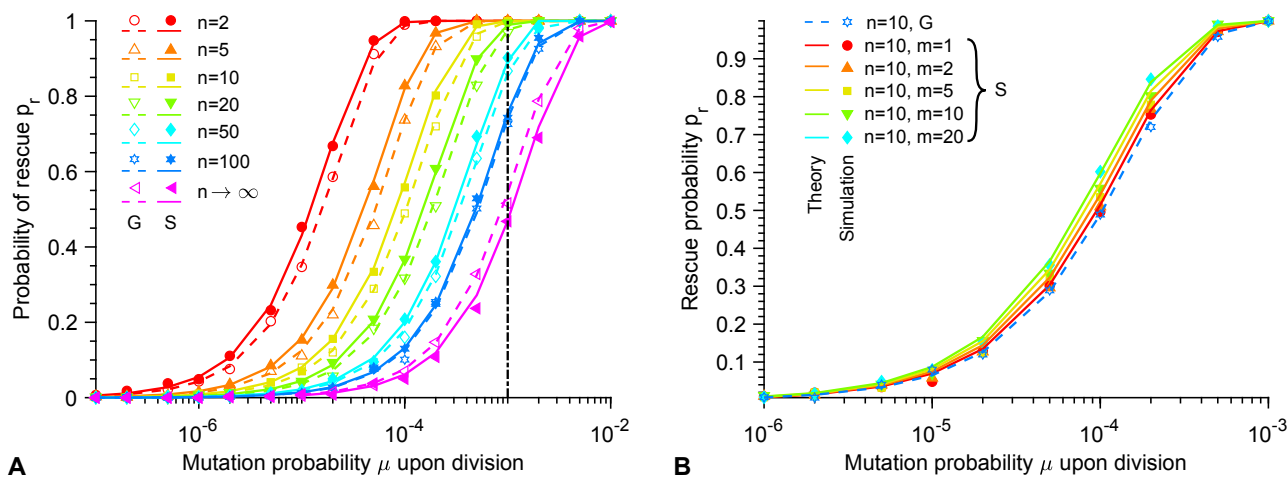}%
	\vspace{0.2cm}
	\caption{{\bf Rescue probability.} {\bf A.} Rescue probability $p_{\mbox{\scriptsize r}}$ of a W population in a deteriorating environment by G or S mutants, versus mutation probability $\mu$ upon division. Different Hill coefficients $n$ characterizing the steepness of the environment deterioration (see Eq.~\ref{HE_eq}) are considered. Here, S mutants satisfy $m=n$, i.e. they have the same sensitivity to the environment as W organisms (see Eq.~\ref{fS}). Vertical dash-dotted line: $K\mu=1$.  {\bf B.} Rescue probability $p_{\mbox{\scriptsize r}}$ by different types of mutants versus mutation probability $\mu$ upon division. A fixed Hill coefficient $n=10$ characterizing the decay of $f_W$ (see Eq.~\ref{HE_eq}) is chosen, but G mutants and S mutants with different Hill coefficients $m$ (see Eq.~\ref{fS}) are considered. In both panels, markers correspond to averages over $10^4$ replicate stochastic simulations (``Simulation''). Dashed and solid lines correspond to numerical integrations of Eq. \ref{paf} (``Theory'') for G and S mutants, respectively. Parameter values: $g_W=g_G=g_S=0.1$, $K=10^3$, $N_W^0=10$ and $\theta=10^3$. }%
	\label{HE_paf}%
\end{figure}

Fig. \ref{HE_paf} shows the rescue probability $p_r$ versus the mutation probability $\mu$ at each division. It demonstrates a very good agreement between our analytical prediction in Eq.~\ref{paf} and results from our stochastic simulations (see Supporting Information, section \ref{Appearance and fixation probability of mutants}). We observe a sigmoidal increase of $p_r$ as $\mu$ increases, with a transition between a small-$\mu$ regime where the population almost certainly goes extinct and a large-$\mu$ regime where it is almost certainly rescued by adaptive mutants. Fig. \ref{HE_paf}A further shows that this transition is strongly impacted by the rapidity of the environment degradation, which is modeled via the Hill coefficient $n$ (see Eq.~\ref{HE_eq}). Specifically, the faster the environment degradation, the bleaker the prospect is for the population, and the larger $\mu$ becomes necessary to allow its rescue. This is related to the rapidity of extinction of the W population in the absence of mutations: for small $n$, the population decay is slower, allowing a larger window of opportunity for mutants to appear and to be selected (see Fig.~\ref{HE_NA}). Increasing $n$ does not substantially affect the steepness of $p_r$, but rather shifts the transition between small and large $p_r$ toward larger $\mu$, because the associated faster decay of the W population mainly decreases the total number of mutants that appear (see Fig.~\ref{NMFig}), with little impact on their fixation probabilities at the end of the process (see Figs.~\ref{HE_pfix}A and ~\ref{pfix_rescaled}). Note that our prediction in Eq.~\ref{paf} is valid far beyond the weak-mutation regime $K\mu\ll1$. While our assumption that $N_W(t)\gg N_M(t)$ when the rescuing mutant arises can fail for $K\mu\gg 1$, rescue is almost certain as this regime is reached. In the limit $n\rightarrow\infty$ of an instantaneous environment degradation, discussed in detail in section \ref{Approx} of the Supporting Information, the transition from large to small $p_r$ occurs for $K\mu\approx1$ (see Fig. \ref{HE_paf}A and Fig.~\ref{Approxpaf}A). Indeed, preexisting mutations then become necessary to population rescue, as no division occurs after the abrupt environment transition. In section \ref{Approx_pr} of the Supporting Information, we further show that Eq.~\ref{paf} generalizes the predictions in our previous work~\cite{Marrec20} regarding the probability of extinction of a microbial population subjected to abrupt additions of antimicrobial, beyond the weak-mutation regime $K\mu\ll1$ (see Fig.~\ref{Approxpaf}B). 

In Fig. \ref{HE_paf}A, we also compare G mutants and S mutants satisfying $m=n$ (see Eq.~\ref{fS}) for each $n$, and we find that S mutants are slightly more successful at rescuing the population than G mutants unless $n$ is very large. This is because S mutants that occur for $t>\theta$ have a larger selective advantage than G mutants and thus a larger fixation probability (see Fig.~\ref{HE_pfix}A). Note that for very steep environment changes, the situation reverses (see Figs.~\ref{HE_paf}A and~\ref{Approxpfix}), because the decay of the W population is so fast that mutants occurring for $t<\theta$ are more likely to be the ones that rescue the population. Consistently, Fig. \ref{HE_paf}B further shows that specialists with a larger Hill coefficient $m$, such that fitness increases more steeply during the environment transition (see Eq.~\ref{fS}), are slightly more efficient at rescuing the population. The impact of $n$ on the rescue probability is stronger than that of $m$, because $n$ controls the rapidity of the decay of the wild-type population, which directly impacts the number of mutants that appear during this decay (see Fig.~\ref{NMFig}).

Apart from the detailed differences we just described, Fig.~\ref{HE_paf} demonstrates that the mutant type affects rescue probability quite little. In section~\ref{antibio} of the Supporting Information, we consider yet another mutant type, aiming to model antimicrobial resistance evolution, and we find that our results are also qualitatively robust to this variant. Overall, the key ingredients are that wild-type organisms are doomed to extinction in the absence of mutants, while mutants are fit enough in the final environment to be able to grow and rescue the population. If this holds, the detailed time evolution of mutant fitness matters little. 

\newpage

\subsection*{Time of appearance of the mutants that fix}

\begin{figure}[h!]%
	\centering
	\includegraphics[width=\textwidth]{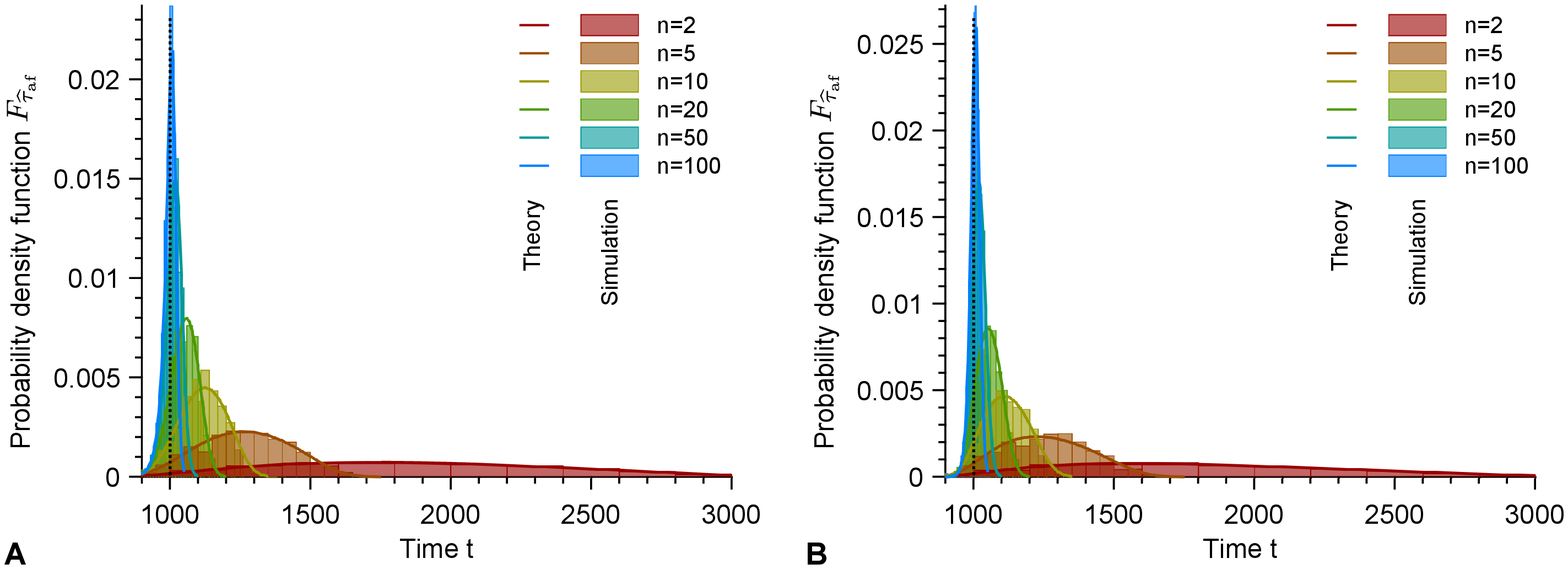}%
	\vspace{0.2cm}
	\caption{{\bf Probability density function of the time of appearance of the mutants that fix.} Probability density function $F_{\widehat{\tau}_{\mathrm{af}}}$ of the time $\widehat{\tau}_{\mathrm{af}}$ of appearance of a mutant that fixes versus time $t$, for different Hill coefficients $n$. Results for the generalist (G) and specialist (S) mutants are shown in panels {\bf A} and {\bf B}, respectively. Vertical dotted line: $t=\theta$. Histograms are computed over $10^3$ replicate stochastic simulations (``Simulation''). Solid lines correspond to numerical integrations of Eq.~\ref{Ftauaf} (``Theory''). Parameter values: $g_W=g_G=g_S=0.1$, $K=10^3$, $\theta=10^3$, $\mu=10^{-5}$, $n=m=5$, and $N_W^0=10$.
	}%
	\label{PDF_tauaf}%
\end{figure}

The fixation probability of a mutant strongly depends on the time at which it appears during the environment degradation (see Fig.~\ref{HE_pfix}). But when do the mutants that fix and rescue the population appear? The probability density function $F_{\widehat{\tau}_{\mbox{\scriptsize af}}}$ of the time $\widehat{\tau}_{\mbox{\scriptsize af}}$ of appearance of a mutant that fixes can be obtained from $p_{\mbox{\scriptsize af}}$ (see Eq.~\ref{dpaf1} and below) through $F_{\widehat{\tau}_{\mbox{\scriptsize af}}}=(1/p_{\mbox{\scriptsize r}})\mbox{d}p_{\mbox{\scriptsize af}}/\mbox{d}t$, where normalization is ensured by $1/p_{\mbox{\scriptsize r}}$ (we focus on cases where rescue occurs). Indeed, $p_{\mbox{\scriptsize af}}(t)/p_{\mbox{\scriptsize r}}$ is the cumulative distribution function of $\widehat{\tau}_{\mbox{\scriptsize af}}$. Thus,
\begin{equation}
F_{\widehat{\tau}_{\mbox{\scriptsize af}}}(t)=\frac{\mu}{p_{\mbox{\scriptsize r}}}\, p_{\mbox{\scriptsize fix}}(t)N_W(t) f_W(t) \left(1-\frac{N_W(t)}{K}\right)\exp({-\Sigma(t)})\mbox{ },
\label{Ftauaf}
\end{equation}
where
\begin{equation}
\Sigma(t)=\mu\int_0^{t}p_{\mbox{\scriptsize fix}}(u)N_W(u)f_W(u)\left(1-\frac{N_W(u)}{K}\right) \mbox{d}u \mbox{ }.
\label{thesigma}
\end{equation}
$F_{\widehat{\tau}_{\mbox{\scriptsize af}}}$ is shown in Fig.~\ref{PDF_tauaf} for different Hill coefficients $n$ characterizing the steepness of the environment deterioration. It illustrates that rescuing mutants tend to appear later as $n$ is decreased, because the decay of the W population is slower in these cases. 

\begin{figure}[h!]%
	\centering
	\includegraphics[width=\textwidth]{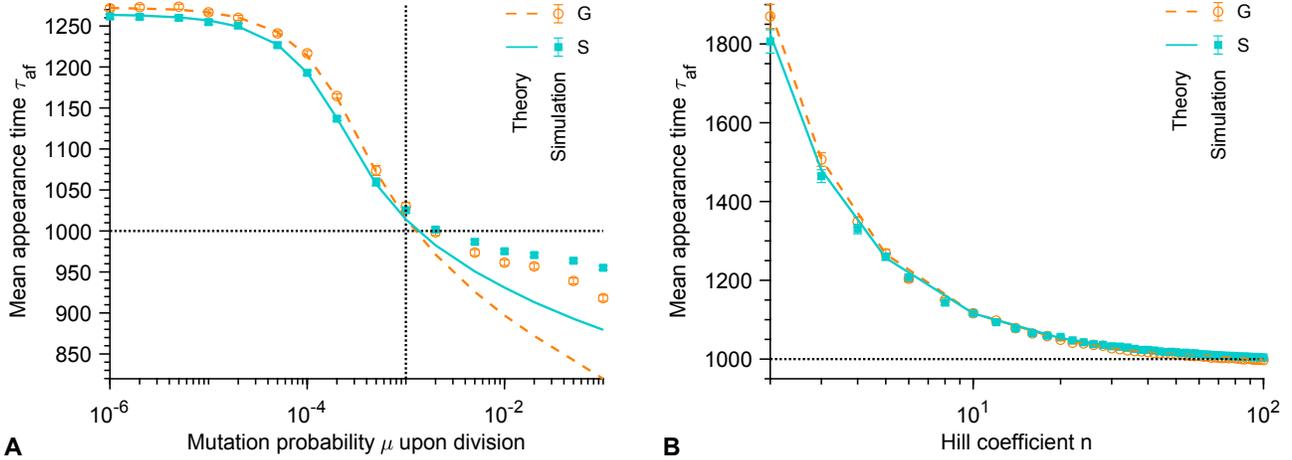}%
	\vspace{0.2cm}
	\caption{{\bf Time of appearance of the mutants that fix.} {\bf A.} Average time $\tau_{\mbox{\scriptsize af}}$ of appearance of a G or S mutant that fixes versus mutation probability $\mu$ upon division. The Hill coefficient characterizing the steepness of the environment deterioration (see Eq.~\ref{HE_eq}) is $n=5$. Vertical dotted line: $K\mu=1$.  {\bf B.} Average time $\tau_{\mbox{\scriptsize af}}$ of appearance of a G or S mutant that fixes versus Hill coefficient $n$. The mutation probability upon division is $\mu=10^{-5}$. 
		In both panels, markers correspond to averages over $10^3-10^4$ replicate stochastic simulations (``Simulation''). Dashed and solid lines correspond to numerical integrations of Eq. \ref{tauaf} (``Theory'') for G and S mutants, respectively. Parameter values: $g_W=g_G=g_S=0.1$, $K=10^3$, $N_W^0=10$ and $\theta=10^3$. Horizontal dotted lines: $\tau_{\mbox{\scriptsize af}}=\theta$.
	}%
	\label{HE_tauaf}%
\end{figure}

Eq. \ref{Ftauaf} allows to express the average time $\tau_{\mbox{\scriptsize af}}=\langle \widehat{\tau}_{\mbox{\scriptsize af}} \rangle$ of appearance of the mutants that fix:
\begin{equation}
\tau_{\mbox{\scriptsize af}}=\int_0^{\infty}tF_{\tilde{\tau}_{\mbox{\scriptsize af}}}(t)\mbox{d}t=\frac{\mu}{p_{\mbox{\scriptsize r}}}\int_0^\infty t\,p_{\mbox{\scriptsize fix}}(t)N_W(t) f_W(t) \left(1-\frac{N_W(t)}{K}\right)\exp({-\Sigma(t)})\,\mbox{d}t \mbox{ }.
\label{tauaf}
\end{equation}
Fig. \ref{HE_tauaf} shows the average time $\tau_{\mbox{\scriptsize af}}$ of appearance of the mutants that fix, and demonstrates a very good agreement between our analytical prediction in Eq.~\ref{tauaf} and the results of our stochastic simulations in the weak-to-moderate mutation regime $K\mu\lesssim1$. (Recall that our calculations assume that $N_W(t)\gg N_M(t)$ when the rescuing mutant appears, which can fail when $K\mu$ is large.) Fig. \ref{HE_tauaf}A shows that $\tau_{\mbox{\scriptsize af}}$ decreases as the mutation probability $\mu$ upon division is increased: this is because more mutants appear for larger $\mu$. In addition, $\tau_{\mbox{\scriptsize af}}$ is larger than the inflection time $\theta$ for $K\mu\lesssim1$, which confirms that the mutants that fix tend to be beneficial ones (see Fig.~\ref{HE_pfix}), and is consistent with the fact that S mutants, which are more beneficial than G mutants for $t>\theta$, are more efficient at rescuing the population (see Fig.~\ref{HE_paf}). Besides, when $\tau_{\mbox{\scriptsize af}}>\theta$, S mutants that fix appear earlier than G mutants that fix: this is also due to their larger selective advantage, and consistently, the opposite holds for $\tau_{\mbox{\scriptsize af}}<\theta$, when G mutants are fitter than S mutants (see Eq.~\ref{HE_eq}). In addition, Fig. \ref{HE_tauaf}B shows that $\tau_{\mbox{\scriptsize af}}$ decreases as the Hill coefficient $n$ which characterizes the steepness of the environment degradation (see Eq.~\ref{HE_eq}) is increased. Indeed, for large $n$, the population gets extinct quickly and rescue needs to occur fast if it occurs at all.

While we have mainly focused on mutants that fix and rescue the population, in section \ref{Extinction time of mutants} of the Supporting Information, we also investigate the mean time to extinction of the lineages of mutants that do not fix. This time is longest for mutants appearing close to the inflection point $\theta$ of the environment transition, which corresponds to the time when the fitness difference between W organisms and mutants is smallest. Intuitively, mutants that are strongly deleterious or beneficial have their fates sealed faster than neutral ones. Furthermore, in the framework of the Moran process (with constant population size and fitnesses), extinction times are longest for neutral mutants~\cite{Ewens79,Teimouri19,Teimouri19b}. While the time to extinction is not crucial to our study of rescue by a single mutation, it can become relevant to more complex processes involving several mutations, e.g. to the crossing of fitness valleys or plateaus~\cite{Weissman09,Bitbol14}.

\newpage

\subsection*{Impact of population size on rescue}
\label{PopSize_sec}

So far, we have discussed population rescue at a given carrying capacity $K$. What is the impact of $K$ on rescue? 

\begin{figure}[h!]%
	\centering
	\includegraphics[width=\textwidth]{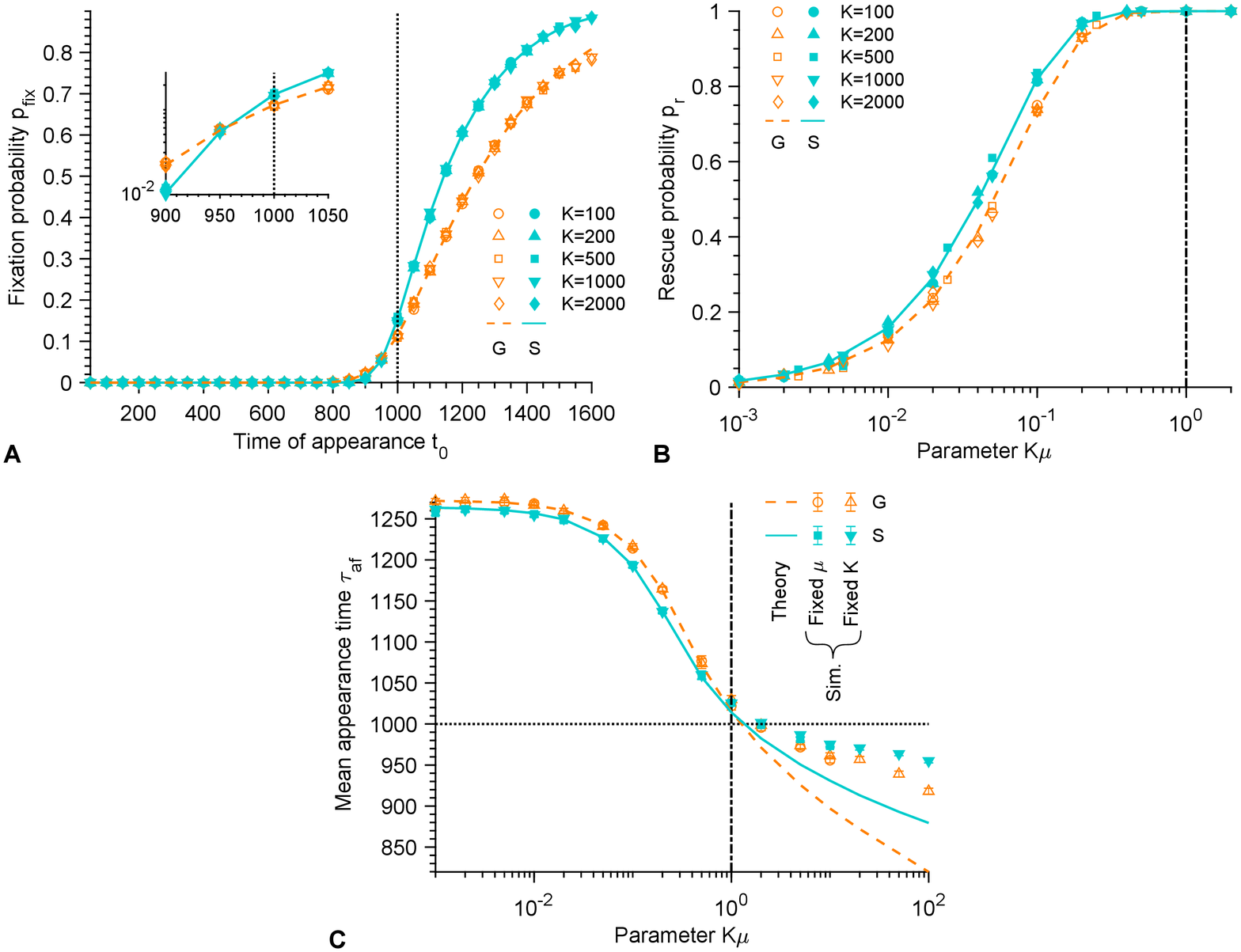} 
	\vspace{0.2cm}
	\caption{{\bf Impact of population size on rescue.} {\bf A.} Fixation probability $p_{\mbox{\scriptsize fix}}$ of G and S mutants versus their time of appearance $t_0$ in the deteriorating environment, for different carrying capacities $K$. Vertical dotted line: $t=\theta$. Main panel: linear scale; inset: semi-logarithmic scale. {\bf B.} Rescue probability $p_r$ of different types of mutants versus the product $K\mu$ of the carrying capacity $K$ and the mutation probability $\mu$ upon division, for different carrying capacities $K$. G mutants and S mutants are considered. Vertical dash-dotted line: $K\mu=1$. {\bf C.} Mean time $\tau_{\mbox{\scriptsize af}}$ of appearance of a G or S mutant that fixes versus $K\mu$. Simulation results are shown both for a fixed mutation probability upon division $\mu=10^{-5}$ and a variable carrying capacity $K$, and for a fixed $K=10^3$ and a variable $\mu$. Horizontal dotted line: $\tau_{\mbox{\scriptsize af}}=\theta$. Vertical dash-dotted line: $K\mu=1$.
		In all panels, the Hill coefficient characterizing the steepness of the environment deterioration (see Eq.~\ref{HE_eq}) is  $n=5$. Furthermore, S mutants satisfy $m=n$, i.e. they have the same sensitivity to the environment as W organisms (see Eq.~\ref{fS}). Markers correspond to averages over $10^3-10^4$ replicate stochastic simulations (``Sim.''). Dashed and solid lines correspond to our analytical predictions (``Theory'') for G and S mutants, respectively. Parameter values: $g_W=g_G=g_S=0.1$, $N_W^0=10$ and $\theta=10^3$.
	}%
	\label{PopSize}%
\end{figure}

First, our analytical expression of the fixation probability $p_{\mbox{\scriptsize fix}}$ of mutants in Eq.~\ref{pfix} depends on $K$ only via the function $\rho$ introduced in Eq.~\ref{rho}. But $\rho$ depends on the number of wild-type microbes $N_W(t)$ and on the carrying capacity $K$ only through the ratio $N_W(t)/K$, whose dynamics is independent from $K$ (see Eq. \ref{dNAdt}). Therefore, $p_{\mbox{\scriptsize fix}}$ is expected to be independent from $K$. Fig.~\ref{PopSize}A confirms that it is the case: the simulation results obtained for different values of $K$ collapse on the same curves. In addition, they are in very good agreement with the predictions from Eq.~\ref{pfix}. Note that Eq.~\ref{text} shows that the mean extinction time of the lineages of mutants that do not fix is also independent from population size, which is confirmed by Fig.~\ref{HE_text}B.

Let us now turn to the rescue probability $p_{\mbox{\scriptsize r}}$. Eqs.~\ref{paf} and~\ref{theSigma} demonstrate that $p_{\mbox{\scriptsize r}}$ depends on population size only via the product $N_W(t) \mu$. Therefore, the relevant parameter is $K\mu$. Fig.~\ref{PopSize}B confirms that $p_{\mbox{\scriptsize r}}$ only depends on $K$ via $K\mu$: the simulation results obtained for different values of $K$ collapse on the same curves when they are plotted as a function of $K\mu$, and feature a good agreement with Eq.~\ref{paf}. For larger $K$, smaller mutation probabilities per division suffice to ensure larger rescue probabilities, because more mutants appear in larger populations, but more precisely, what really matters for rescue is the value of $K\mu$. This finding extends previous results regarding abrupt environment change~\cite{Martin12}. 

Finally, Eqs.~\ref{thesigma} and~\ref{tauaf} show that for the mean time $\tau_{\mbox{\scriptsize af}}$ of appearance of a mutant that fixes, the relevant parameter is also $K\mu$. Fig.~\ref{PopSize}C confirms this: the simulation results obtained by varying $\mu$ at constant $K$ or by varying $K$ at constant $\mu$ collapse when they are plotted as a function of $K\mu$, in good agreement with Eq.~\ref{tauaf}.

Overall, the main quantities that characterize population rescue, namely the rescue probability $p_{\mbox{\scriptsize r}}$ and the mean time $\tau_{\mbox{\scriptsize af}}$ of appearance of a mutant that fixes, are governed by $K\mu$. Hence, the impact of population size and mutation probability is mainly felt through this parameter.

\newpage

\section*{Discussion}

In this paper, we investigated the evolutionary rescue of a microbial population in a gradually deteriorating environment, characterized by a sigmoidal decay down to zero of the fitness of wild-type organisms, with a tunable steepness. The population is thus destined for extinction in the absence of adaptive mutants. We showed that mutants that appear later during the environment deterioration have a higher probability to fix, due to an increase of their intrinsic fitness advantage and to competitive release~\cite{Wargo07,Kouyos14}. However, the decay of the wild-type population also entails that mutants are less likely to appear at such late stages. We demonstrated that the overall rescue probability of the population increases with a sigmoidal shape as the product $K\mu$ of the carrying capacity $K$ and of the mutation probability $\mu$ is increased, which extends previous results regarding abrupt environment change~\cite{Martin12}. In the limit of an instantaneous environment degradation, the increase of rescue probability occurs for $K\mu\approx 1$, as preexisting mutations become necessary for rescue. Importantly, much smaller values of $K\mu$ suffice for rescue if the environment degradation, and thus the population decay, are slower, consistently with previous studies on the rate of fitness decay in the regime of stabilizing selection~\cite{Burger95,Gomulkiewicz09}. We also found that our main conclusions are robust to the exact type of mutant considered (generalist, specialist or modeling antimicrobial resistance evolution), provided that mutants are fit enough in the final environment to be able to rescue the microbial population, which is doomed to extinction in the absence of mutants. We further characterized the rescue process by investigating the average time of appearance of the mutants that do rescue the population, which also depends on the parameter $K\mu$, and the average extinction time of those that do not, which is longest when mutants are almost neutral. 

In all cases, we provided both analytical expressions and stochastic simulation results, and obtained a very good agreement between them. Our analytical expressions were obtained with assumptions that hold in the weak-to-moderate mutation regime $K\mu\lesssim 1$, as we only required the wild-type population to be much larger than the mutant one upon the appearance of the successful mutant lineage. Our methods can be applied to other situations with continuously variable fitnesses and population sizes. Our predictions could be tested in controlled evolution experiments, e.g. in the context of antimicrobial resistance evolution, especially by varying population size and/or by studying strains with different mutation rates.

Overall, our study quantitatively confirms the key impact of the rapidity of environment degradation on the fate of a population, with fast degradation bringing the harshest prospects for population survival. This point confirms and extends previous theoretical results regarding a trait under stabilizing selection with a gradually moving optimum~\cite{Burger95}, as well as experimental~\cite{Lindsey13} and numerical~\cite{Wu14} results in the context of antibiotic resistance. Very large populations can almost always escape extinction because they have a wide range of existing mutants, while smaller ones (or rarely mutating ones, since what matters is $K\mu$) can be rescued by adaptive mutations only if the environment changes slowly enough. The case of not-too-large populations is practically very important because real populations tend to have complex structures~\cite{VanMarle07}, and competition is local, which decreases their effective size, at least on timescales shorter than those of large-scale migrations and/or mixing. Accordingly, an exciting extension would be to consider the impact of spatial structure~\cite{Bitbol14,Nahum15,Cooper15} on evolutionary rescue~\cite{Uecker14,CzupponPreprint} in a gradually deteriorating environment. In cases where one aims to avoid rescue, our results entail that environment changes should be made as fast as possible. For instance, in order to avoid antimicrobial resistance evolution, gradually increasing doses of antimicrobial should be avoided. In addition, our results on the fixation probability of mutants and on the mean time of appearance of mutants that fix could be exploited in evolution experiments, e.g. to t mutagen use to potentially favor the appearance of rescue mutants. The average time to extinction of mutants that do not fix can also matter in practice, as another environment change occurring within this time after their appearance might rescue them. Importantly, here, we have considered rescue by a single mutation. However, more adaptations can be accessible in several mutation steps, and thus, considering rescue in a gradually deteriorating environment in the presence of fitness valleys~\cite{Weinreich05,Weissman09} or on more complete fitness landscapes~\cite{Poelwijk07,Szendro13} would also be very interesting from a theoretical point of view. Studying the interplay between time variability of the environment and spatial heterogeneities would also be interesting in this context, given that static antimicrobial gradients can favor resistance evolution~\cite{Zhang11,Greulich12,Hermsen12,Baym16}, in particular by stepwise accumulation of several mutations.

\section*{Acknowledgments}
LM acknowledges funding by a graduate fellowship from EDPIF.

\newpage

\beginsupplement


\begin{center}
\huge{\textbf{Supporting Information}}
\end{center}

\vspace{1cm}

\tableofcontents

\newpage

\section{Fixation probability of mutants}
\label{probaFix}

\subsection{Derivation}

Here, we present the derivation of the fixation probability $p_{\mbox{\scriptsize fix}}(i_0,t_0)$ of $i_0$ mutants present at time $t_0$~\cite{Uecker11,Alexander12,Parzen}, along similar lines as in~\cite{Uecker11}. We assume that the number of wild-type microorganisms is initially much larger than the number of mutants ($N_W(t_0) \gg i_0$). As explained in the main text,  the selective pressure due to the competition with the wild-type is felt by the mutants through their division rate $f_M(t)[1-N(t)/K]$, and in the initial phase where this competition is important, the total population size N(t) can be approximated by $N(t)\approx N_W(t)$. Thus, competition is felt through the effective mutant fitness $f_M^{\mbox{\scriptsize eff}}(t)=f_M(t)[1-N_W(t)/K]$. In addition, we treat the number of mutants stochastically, but the number $N_W(t)$ of wild-type organisms deterministically (see Eq. \ref{dNAdt} and Fig. \ref{HE_NA}). 

The master equation that describes the evolution of the probability $P(i, t|i_0, t_0)$ of having $i$ mutants at time $t$ knowing that there are $i_0$ mutants at time $t_0$ is given by:
\begin{equation}
\frac{\partial P(i, t|i_0, t_0)}{\partial t}=f_M^{\mbox{\scriptsize eff}}(t)(i-1)P(i-1, t|i_0, t_0)+g_M(i+1)P(i+1, t|i_0, t_0)-(f_M^{\mbox{\scriptsize eff}}(t)+g_M)iP(i, t|i_0, t_0)\mbox{ }.
\label{ME}
\end{equation}  
Eq. \ref{ME} allows to establish the partial differential equation satisfied by the probability generating function $\phi_{i_0, t_0}(z, t)=\sum_{i=0}^{+ \infty}z^iP(i, t|i_0, t_0)$:
\begin{equation}
\frac{\partial \phi_{i_0, t_0}}{\partial t}=(z-1)(f_M^{\mbox{\scriptsize eff}}(t)z-g_M)\frac{\partial \phi_{i_0, t_0}}{\partial z} \mbox{ }.
\label{PGFeq}
\end{equation}
The method of characteristics then yields \cite{Kendall48, Parzen}:
\begin{equation}
\phi_{i_0, t_0}(z, t)=\left[1+\left(\frac{e^{\rho(t)}}{z-1}-\int_{t_0}^tf_M^{\mbox{\scriptsize eff}}(u)e^{\rho(u)}\mbox{d}u\right)^{-1}\right]^{i_0}\mbox{ },
\label{PGF}
\end{equation}
where:
\begin{equation}
\rho(t)=\int_{t_0}^t(g_M-f_M^{\mbox{\scriptsize eff}}(u))\mbox{d}u\mbox{ }.
\end{equation}
Note that $\rho$ depends on the number of wild-type microbes $N_W(t)$ and on the carrying capacity $K$ only through the ratio $N_W(t)/K$, whose dynamics is system size-independent, i.e. independent from $K$ (see Eq. \ref{dNAdt}). 

The probability generating function $\phi_{i_0, t_0}$ allows to calculate the fixation probability $p_{\mbox{\scriptsize fix}}(i_0,t_0)$ of $i_0$ mutants present at time $t_0$, through  $p_{\mbox{\scriptsize fix}}(i_0,t_0)=1-\lim_{t\to\infty}P(0,t|i_0,t_0)=1-\lim_{t\to\infty}\phi_{i_0, t_0}(0,t)$. This yields
\begin{equation}
p_{\mbox{\scriptsize fix}}(i_0, t_0)=1-\left(\frac{g_M\int_{t_0}^{\infty} e^{\rho(t)}\mbox{d}t}{1+g_M\int_{t_0}^{\infty} e^{\rho(t)}\mbox{d}t}\right)^{i_0},
\label{pfix0}
\end{equation}
where we used:
\begin{equation}
\int_{t_0}^t(g_M-f_M^{\mbox{\scriptsize eff}}(u))e^{\rho(u)}\mbox{d}u=e^{\rho(t)}-1 \mbox{ }.
\label{astuce}
\end{equation}
Since $\rho$ does not depend on the carrying capacity $K$, as noted above, this is also true for $p_{\mbox{\scriptsize fix}}$ (see Fig.~\ref{PopSize}A). 

In the main text, we focus on the fixation probability of a single mutant that appears at time $t_0$, and denote it as $p_{\mbox{\scriptsize fix}}(t_0)=p_{\mbox{\scriptsize fix}}(1,t_0)$ (see Eq.~\ref{pfix}, which corresponds to Eq.~\ref{pfix0} with $i_0=1$). 

\subsection{Additional results}

\begin{figure}[h!]%
	\centering
	\includegraphics[width=0.5\textwidth]{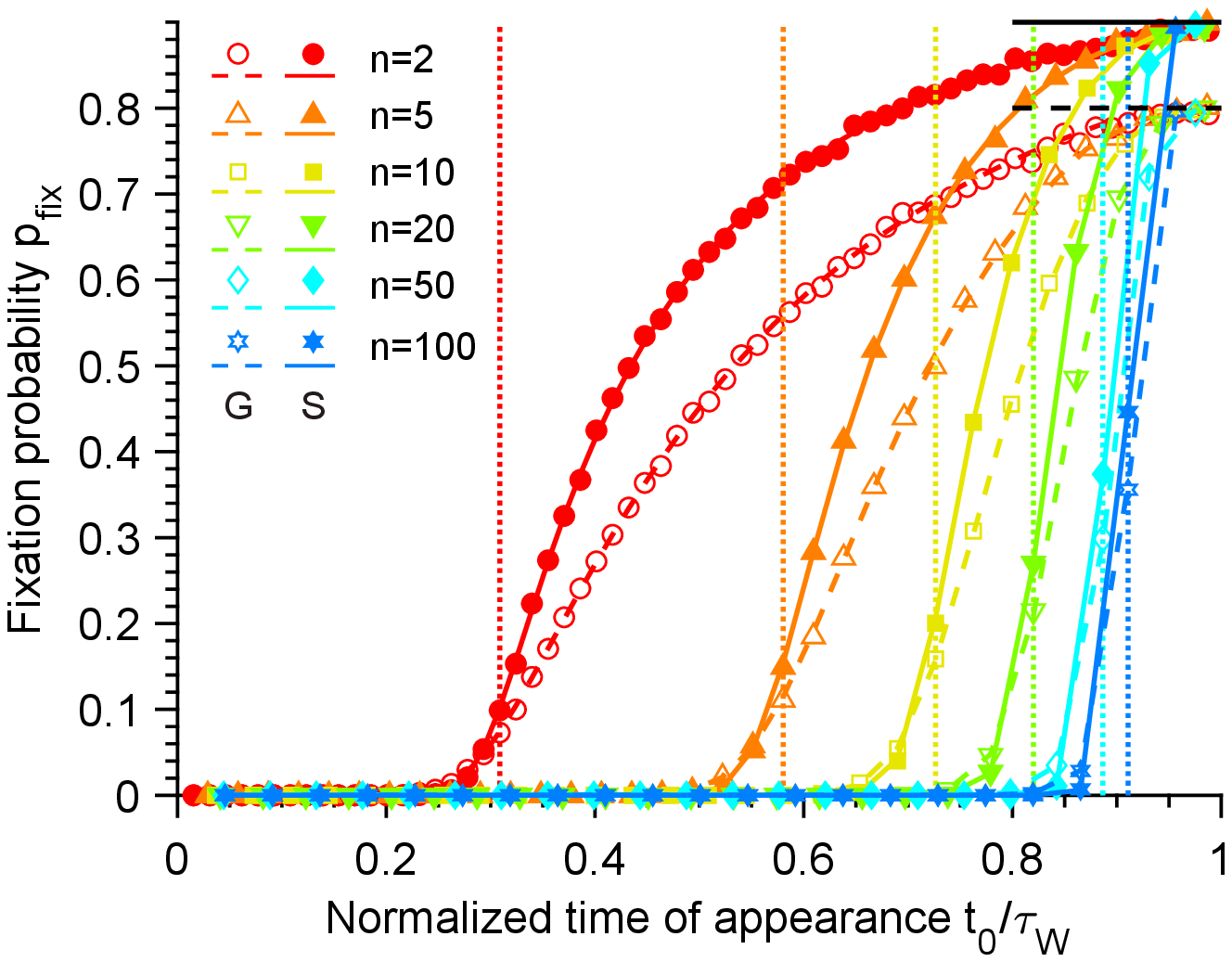}
	\vspace{0.2cm}
	\caption{{\bf Fixation probability of mutants.} Fixation probability $p_{\mbox{\scriptsize fix}}$ of G and S mutants versus their time of appearance $t_0$ in the deteriorating environment, rescaled by the average extinction time $\tau_W$ of the wild-type population for different Hill coefficients $n$ characterizing the steepness of the environment deterioration (see Eq.~\ref{HE_eq}). Here, S mutants satisfy $m=n$, i.e. they have the same sensitivity to the environment as W organisms (see Eq.~\ref{fS}). Horizontal dashed line: $p_{\mbox{\scriptsize fix}}=1-g_G/f_G$. Horizontal solid line: $p_{\mbox{\scriptsize fix}}=1-g_S$. Markers correspond to averages over $10^4$ replicate stochastic simulations. Dashed and solid lines correspond to numerical integrations of Eq. \ref{pfix} for G and S mutants, respectively. Parameter values: $g_W=g_G=g_S=0.1$, $K=10^3$, $N_W^0=10$ and $\theta=10^3$. Vertical dotted lines: $t_0=\theta$. Main panels: linear scale; insets: semi-logarithmic scale. Same data as in Fig.~\ref{HE_pfix}A. }%
	\label{pfix_rescaled}%
\end{figure}

Fig.~\ref{pfix_rescaled} shows the same data as in Fig.~\ref{HE_pfix}A for the fixation probability $p_{\mbox{\scriptsize fix}}$ of G and S mutants versus their time of appearance $t_0$ in the deteriorating environment. However, here, $t_0$ is rescaled by the average extinction time $\tau_W$ of the wild-type population in the absence of mutation (see Fig.~\ref{HE_NA}). This rescaling illustrates the convergence of $p_{\mbox{\scriptsize fix}}$ toward asymptotes independent of $n$ as $\tau_W$ is approached. These asymptotes correspond to the extinction probabilities of mutants that exist in the absence of competition: mutants fix unless their lineage undergoes rapid stochastic extinction.

\newpage

\section{Application to different types of mutants}

\subsection{Antimicrobial resistance evolution}
\label{antibio}

\begin{figure}[h!]
	\centering
	\includegraphics[width=0.5\textwidth]{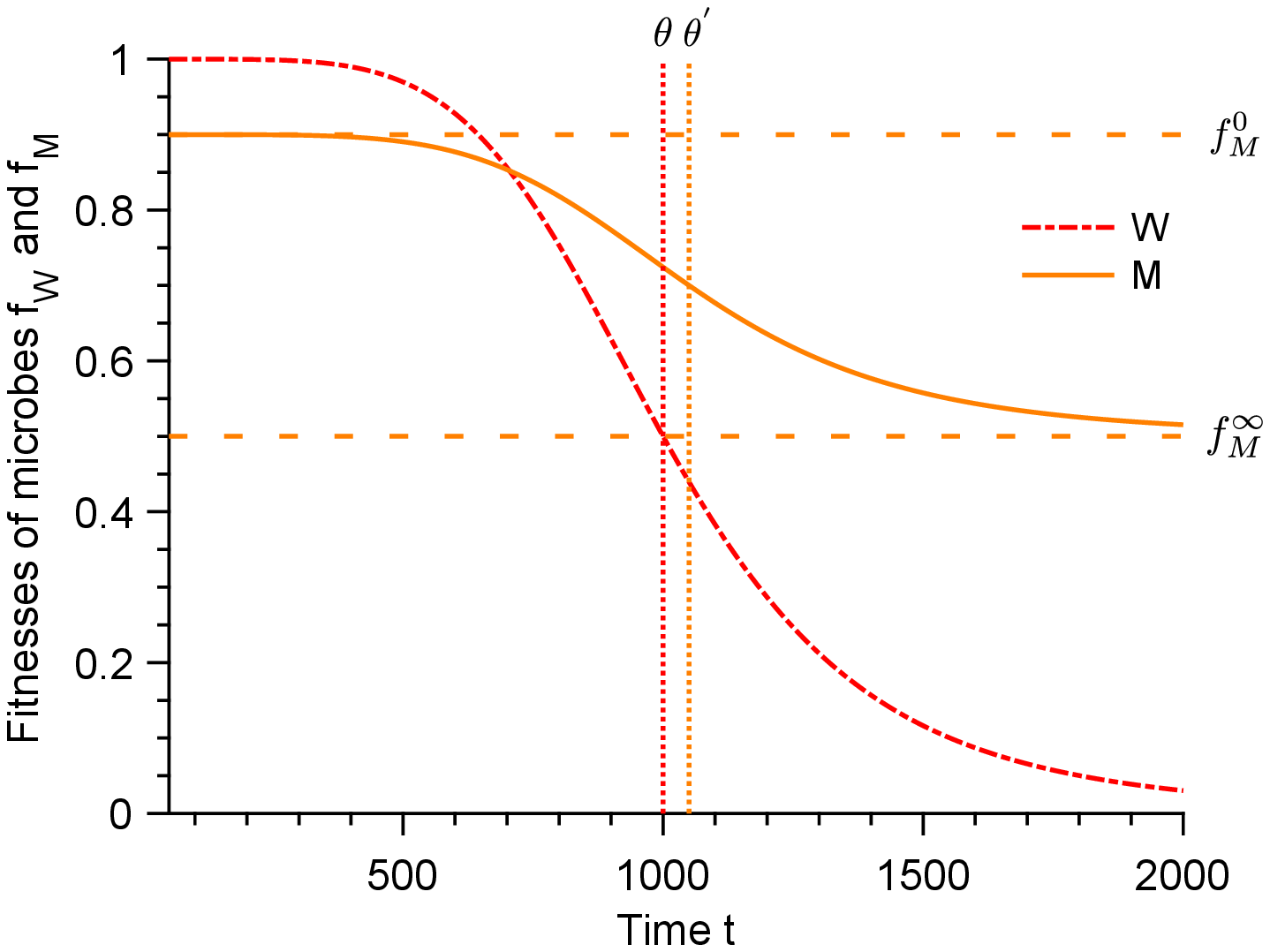}
	\vspace{0.2cm}
	\caption{{\bf Fitnesses of the wild-type and mutant microbes in a model of antimicrobial resistance evolution.} Fitnesses $f_W$ and $f_M$ of the wild-type sensitive microorganisms (W) and resistant mutants (M) versus time $t$ (see Eqs.~\ref{HE_eq} and~\ref{fMres}). Parameter values: $n=5$, $\theta=1000$, $\theta'=1050$, $f_M^0=0.9$ and $f_M^\infty=0.5$. Vertical dotted lines: $t=\theta$ and $t=\theta'$. Horizontal dashed lines: $f_M^0$ and $f_M^\infty$.}%
	\label{FitnessR}%
\end{figure}

An important application of the study of evolutionary rescue regards antimicrobial resistance evolution, where rescue of the microbial population corresponds to the fixation of resistance. In line with our model comprising two types of individuals, let us consider sensitive wild type microbes W, and resistant mutants M. Furthermore, because we consider variable fitnesses and constant death rates (as throughout this work), we here model the effect of biostatic antimicrobials, and not biocidal ones. However, our model could easily be extended to the biocidal case. Let us assume that the concentration of antimicrobial gradually increases from 0 to some value which is above the minimum inhibitory concentration (MIC) of the sensitive strain but below the MIC of the resistant strain. Then, appearance and fixation of resistant mutants is necessary for the microbial population to be rescued. 
Let us model the fitness of resistant mutants M by
\begin{equation}
f_M(t)=\frac{f_M^0-f_M^\infty}{1+(t/\theta')^n}+f_M^\infty\,,
\label{fMres}
\end{equation}
which is equal to $f_M^0$ for $t=0$ and tends to $f_M^\infty$ for $t\rightarrow\infty$ (see Fig.~\ref{FitnessR}). Because antimicrobial resistance often comes with a fitness cost in the absence of drug~\cite{Borman96,Andersson10,zurWiesch11}, we will consider $f_M^0<1$. Since the final concentration is assumed to be above the mutant MIC, we have $f_M^\infty>g_M$, which ensures that a resistant population does not go extinct deterministically in the final environment. We further allow for the inflection point $\theta'$ to be different from that of $f_W$, which is $\theta$ (see Eq.~\ref{HE_eq}), so that $\theta'>\theta$ may reflect the fact that M is less sensitive to the environment change than W. Indeed, compared to that of sensitive microorganisms, the dose-response curve of resistant microorganisms is usually shifted towards higher drug concentrations~\cite{Gullberg11,Yu18}. Note that the functional forms taken for $f_W$ and $f_M$ (see Eqs.~\ref{HE_eq} and~\ref{fMres}) are realistic e.g. in the case of a linear drug concentration increase with time, given the usual pharmacodynamics of antibiotics~\cite{Regoes04}.

\newpage

\begin{figure}[h!]
	\centering
	\includegraphics[width=\textwidth]{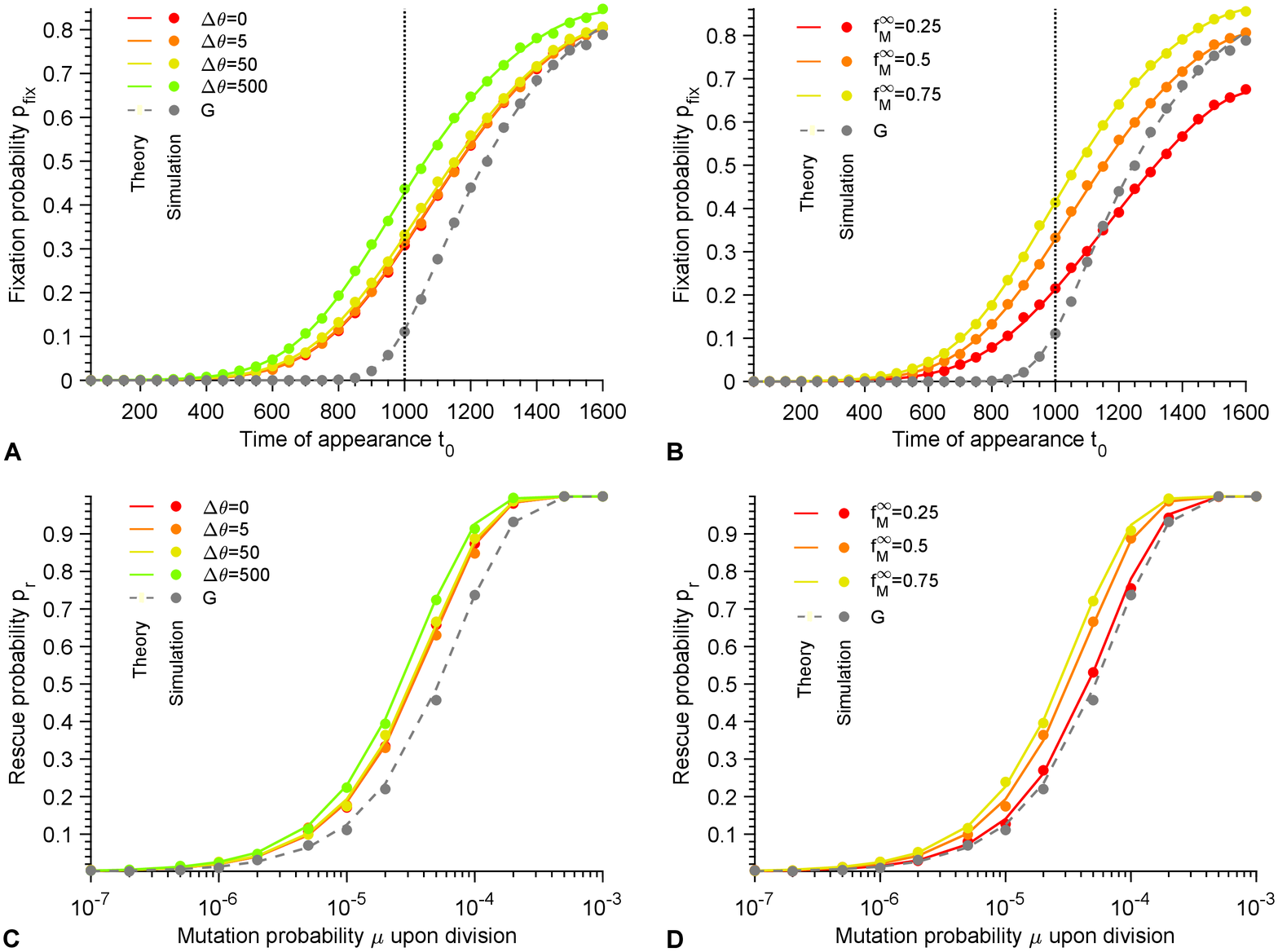}%
	\vspace{0.2cm}
	\caption{{\bf Fixation probability of mutants and probability of rescue in a model of antimicrobial resistance evolution.} {\bf A.} Fixation probability $p_{\mbox{\scriptsize fix}}$ as a function of the time of appearance of the mutants $t_0$ for mutants M with different values of $\Delta\theta=\theta'-\theta$ and $f_M^\infty=0.5$ (see Eqs.~\ref{HE_eq} and~\ref{fMres}) and for generalist (G) mutants with $f_G=0.5$. Vertical dotted line: $t_0=\theta$. {\bf B.} Same as in panel A, but with $\Delta\theta=50$ and different values of $f_M^\infty$. {\bf C.} Rescue probability $p_r$ as a function of the mutation probability $\mu$ upon division for mutants M with different values of $\Delta\theta=\theta'-\theta$ and $f_M^\infty=0.5$ (see Eqs.~\ref{HE_eq} and~\ref{fMres}) and for generalist (G) mutants with $f_G=0.5$, as in panel A. {\bf D.} Same as in panel C, but with $\Delta\theta=50$ and different values of $f_M^\infty$, as in panel B. In all panels, markers correspond to the average over $10^3-10^4$ replicate stochastic simulations, and dashed curves correspond to our analytical predictions. Parameter values: $g_W=g_M=g_G=0.1$, $f_M^0=0.9$, $K=10^3$, $N_W^0=10$, $n=5$ and $\theta=10^3$.}%
	\label{ResultsRes}%
\end{figure}

Fig.~\ref{ResultsRes} shows the results obtained for rescue within this model, and a comparison to the generalist (G) mutant with $f_G=0.5$ studied in the main text. The agreement between our numerical simulations and our analytical predictions is very good. Larger values of $\theta'$ or of $f_M^\infty$ increase the mutant fixation probability $p_{\mbox{\scriptsize fix}}$ and the rescue probability $p_r$, consistently with the fact that they lead to higher mutant fitnesses. Despite minor quantitative differences associated to these parameter values, the rescue probability behaves qualitatively in the same way in this model as with the generalist mutant and as with the specialist mutant studied in the main text. This illustrates the generality of our findings with respect to the exact mutant fitness form, as long as the mutant is able to grow in the new environment and rescue the population.

\clearpage

\subsection{Additional results for various generalist mutants}\label{MoreResults}

\begin{figure}[h!]
	\centering
	\includegraphics[width=\textwidth]{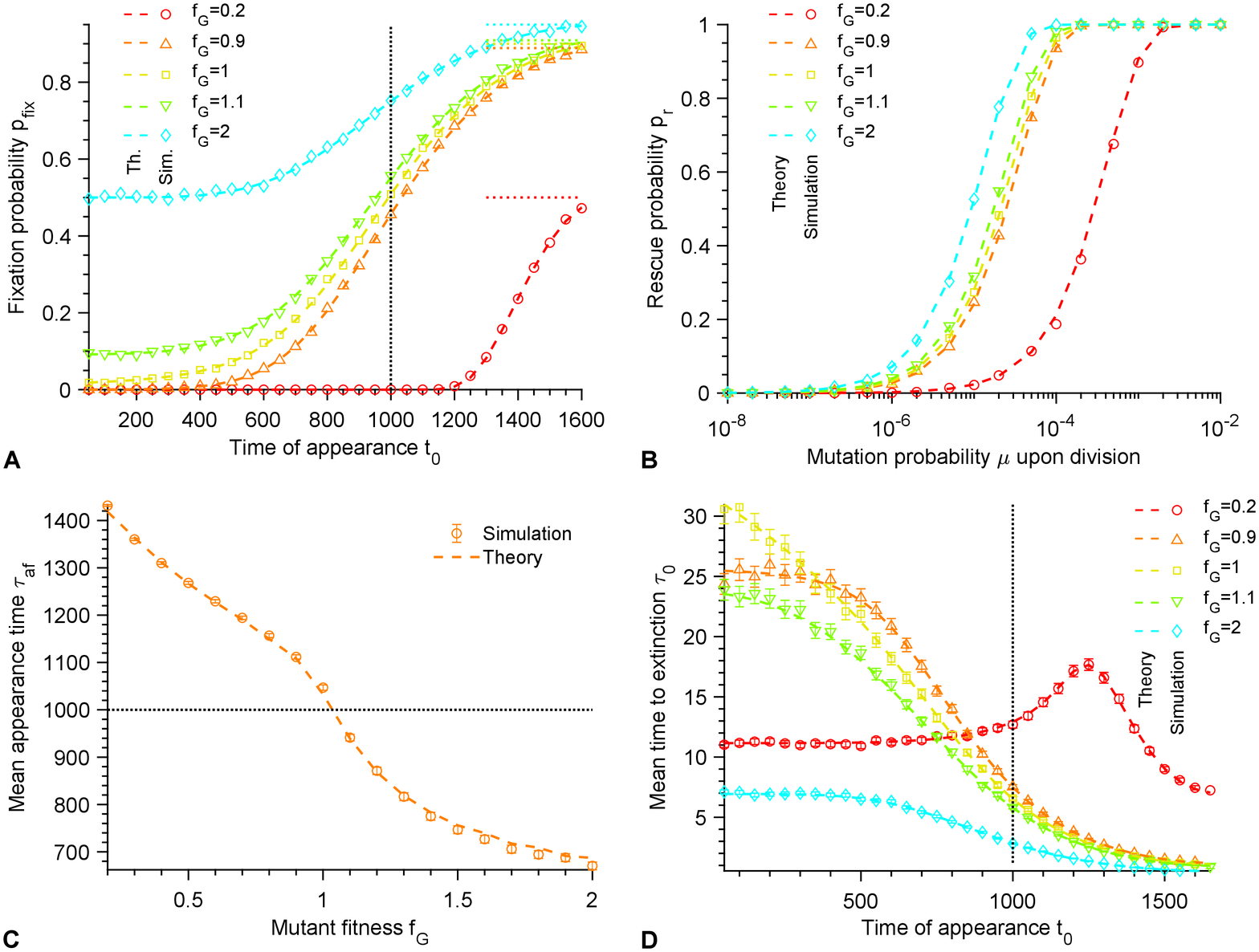}%
	\vspace{0.2cm}
	\caption{{\bf Additional results for generalist mutants.} {\bf A.} Fixation probability $p_{\mbox{\scriptsize fix}}$ as a function of the time of appearance of the mutants $t_0$ for different fitnesses $f_G$ of G mutants (in the rest of the paper, $f_G=0.5$). Vertical dotted line: $t_0=\theta$. Horizontal dotted lines: $p_{\mbox{\scriptsize fix}}=1-g_G/f_G$. {\bf B.} Rescue probability $p_r$ as a function of the mutation probability $\mu$ upon division for different fitnesses $f_G$. {\bf C.} Mean appearance time $\tau_{\mbox{\scriptsize af}}$ of a mutant that fixes as a function of the fitness $f_G$ for the mutation probability upon division $\mu=10^{-5}$. Vertical dotted line: $\tau_{\mbox{\scriptsize af}}=\theta$. {\bf D.} Mean time to extinction $\tau_0$ as a function of the time of appearance of the mutants $t_0$ for different fitnesses $f_G$. Vertical dotted line: $t_0=\theta$. In all panels, markers correspond to the average over $10^3-10^4$ replicate stochastic simulations, error bars (in panels C and D, often smaller than markers) are 95\% confidence intervals and dashed curves correspond to our analytical predictions. Parameter values: $g_W=g_G=0.1$, $K=10^3$, $N_W^0=10$, $n=5$ and $\theta=10^3$.}%
	\label{MoreG}%
\end{figure}

In the main text, we consider generalist (G) mutants with fitness $f_G=0.5$, corresponding to the case of specialist (S) mutants with $m=0$ (see Eq.~\ref{fS}). Fig.~\ref{MoreG} shows results obtained for various values of $f_G$ that satisfy $f_G>g_G$, ensuring that the mutant can grow and rescue the population. Mutant fixation and rescue are more difficult for smaller values of $f_G$, but the overall behavior remains similar and is well described by our analytical predictions.

\clearpage

\section{Robustness of the results to different initial conditions}

\begin{figure}[h!]%
	\centering
	\includegraphics[width=\textwidth]{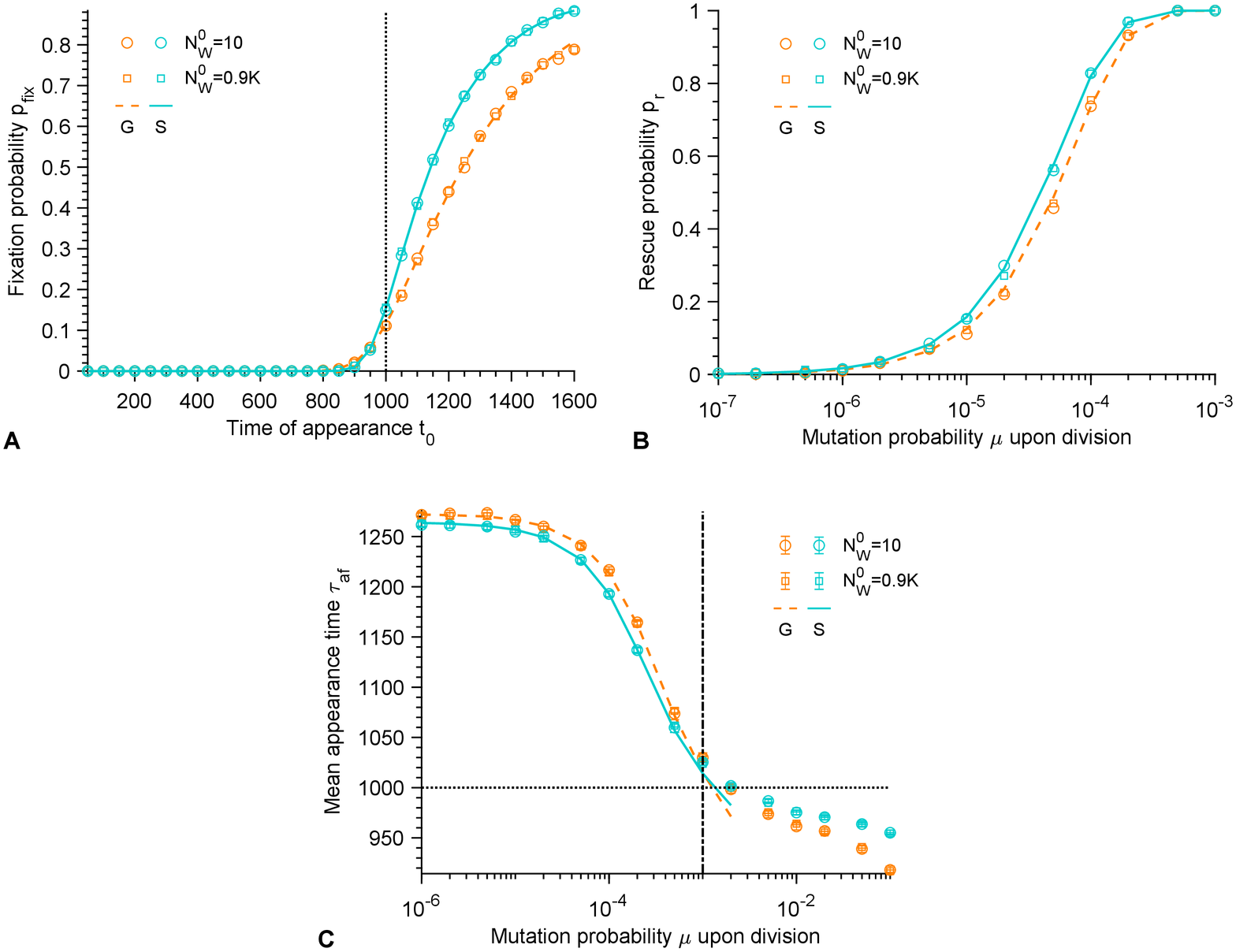}%
	\vspace{0.2cm}
	\caption{{\bf Impact of the initial number $N_W^0$ of wild-type organisms on rescue.} {\bf A.} Fixation probability $p_{\mbox{\scriptsize fix}}$ of G and S mutants versus their time of appearance $t_0$ in the deteriorating environment, for $N_W^0=10$ and $N_W^0=0.9 K$. Vertical dotted line: $t_0=\theta$. {\bf B.} Rescue probability $p_r$ of different types of mutants versus the mutation probability $\mu$ upon division, for $N_W^0=10$ and $N_W^0=0.9 K$. G mutants and S mutants are considered. {\bf C.} Mean time $\tau_{\mbox{\scriptsize af}}$ of appearance of a G or S mutant that fixes versus $\mu$, for $N_W^0=10$ and $N_W^0=0.9 K$. Horizontal dotted line: $\tau_{\mbox{\scriptsize af}}=\theta$. Vertical dash-dotted line: $K\mu=1$. In all panels, the Hill coefficient characterizing the steepness of the environment deterioration (see Eq.~\ref{HE_eq}) is  $n=5$. Furthermore, S mutants satisfy $m=n$, i.e. they have the same sensitivity to the environment as W organisms (see Eq.~\ref{fS}). Markers correspond to averages over $10^3-10^4$ replicate stochastic simulations. Dashed and solid lines correspond to our analytical predictions for G and S mutants, respectively. Parameter values: $g_W=g_G=g_S=0.1$, $K=10^3$ and $\theta=10^3$.
	}%
	\label{Robust}%
\end{figure}

In Fig.~\ref{Robust}, we show that our results are robust to varying $N_W^0$ as long as it is not very small, since starting with $N_{W}^0=10$ (as is done throughout) gives the same results as starting with $N_{W}^0=K[1-g_W/f_W(0)]=0.9K$, which corresponds to the stationary population size in the initial environment within a deterministic description (see Eq.~\ref{dNAdt}). 

\clearpage

\section{Additional results regarding the appearance of mutants}

\subsection{Appearance of mutants during the environment deterioration}

\newpage

\subsection{Time of appearance of the mutants that fix}

\newpage

\section{Results for the impact of population size on rescue}

\clearpage

\section{Extinction time of mutants that do not fix}
\label{Extinction time of mutants}

In the case where the mutant that appears does not fix, how long does its lineage take to go extinct? As for the fixation probability $p_{\mbox{\scriptsize fix}}$, the time of extinction of a mutant will depend on its time of appearance $t_0$. The average time of extinction is the average of the first-passage time $\widehat{\tau}^{'}_0$ to the state $i=0$ where $i$ denotes the number of mutants. Then, we can compute the probability $dp(\widehat{\tau}^{'}_0\in[t, t+\mbox{d}t]\,|\,i_0, t_0)$ that $\widehat{\tau}^{'}_0$ belongs to the interval $[t, t+\mbox{d}t]$, provided that the initial number of mutants is $i_0$ at time $t_0$:
\begin{equation}
dp(\widehat{\tau}^{'}_0\in[t, t+\mbox{d}t]\,|\,i_0, t_0)= P(0, t+dt|0, \infty; i_0,t_0)-P(0, t|0, \infty; i_0,t_0)\mbox{ },
\end{equation}
where $P(0, t|0, \infty; i_0,t_0)$ is the probability to have 0 mutant at time $t$, provided that the initial number of mutants is $i_0$ at time $t_0$ and the final number is $i_{\infty}=0$, corresponding to extinction.
Using Bayes' theorem and the Markov property yields 
\begin{equation}
P(0, t|0, \infty; i_0,t_0)=\frac{P(0,t|i_0,t_0)\,P(0, \infty|0, t; i_0, t_0)}{P(0,\infty|i_0,t_0)}=\frac{P(0,t|i_0,t_0)\,(1-p_{\mbox{\scriptsize fix}}(0,t))}{1-p_{\mbox{\scriptsize fix}}(i_0,t_0)}=\frac{P(0,t|i_0,t_0)}{1-p_{\mbox{\scriptsize fix}}(i_0,t_0)}\,,
\end{equation} 
where we have employed $p_{\mbox{\scriptsize fix}}(0,t)=0$, as having 0 mutant is an absorbing state of the system. Thus,
\begin{equation}
dp(\widehat{\tau}^{'}_0\in[t, t+\mbox{d}t]\,|\,i_0, t_0)=\frac{P(0,t+\mbox{d}t|i_0,t_0)-P(0,t|i_0,t_0)}{1-p_{\mbox{\scriptsize fix}}(i_0,t_0)}=\frac{1}{1-p_{\mbox{\scriptsize fix}}(i_0,t_0)}\frac{\mbox{d}P(0, t|i_0,t_0)}{\mbox{d}t}\mbox{d}t \mbox{ }.
\label{p}
\end{equation}

We can now express the mean time of extinction $\tau^{'}_0=\langle \widehat{\tau}^{'}_0 \rangle$ of a mutant that appeared at $t_0$ using Eq.~\ref{p} as
\begin{equation}
\tau_0^{'}=\int_{t_0}^{\infty}t\,dp(\widehat{\tau}^{'}_0\in[t, t+\mbox{d}t]\,|\,i_0, t_0)=\frac{1}{1-p_{\mbox{\scriptsize fix}}(i_0,t_0)}\int_{t_0}^{\infty}t\frac{\mbox{d}P(0, t|i_0,t_0)}{\mbox{d}t}\mbox{d}t \mbox{ }.
\end{equation}
The previous equation can be rewritten using the probability generating function $\phi_{i_0, t_0}(z, t)=\sum_{i=0}^{+ \infty}z^iP(i, t|i_0, t_0)$ by noting that $P(0, t|i_0, t_0)=\phi_{i_0,t_0}(0,t)$: 
\begin{equation}
\tau_0^{'}=\frac{1}{1-p_{\mbox{\scriptsize fix}}(i_0,t_0)}\int_{t_0}^{\infty}t\,\frac{\partial \phi_{i_0,t_0}}{\partial t}(0,t)\,\mbox{d}t \mbox{ }.
\end{equation}
Using Eqs. \ref{PGF} and \ref{astuce} and introducing $\Lambda(t)=g_M\int_{t_0}^{t}e^{\rho(u)}\mbox{d}u$ then yields
\begin{equation}
\tau_0^{'}=\frac{i_0 g_M}{1-p_{\mbox{\scriptsize fix}}(i_0,t_0)}\int_{t_0}^{\infty}te^{\rho(t)}\frac{\Lambda^{i_0-1}(t)}{(1+\Lambda(t))^{i_0+1}}\mbox{d}t \mbox{ }.
\label{text}
\end{equation}
Numerical integration of Eq. \ref{text} is discussed in section \ref{Numerical resolution details} below.

Fig. \ref{HE_text} shows the average lifetime $\tau_0=\tau_0^{'}-t_0$, or time to extinction, of the lineage of a single mutant ($i_0=1$) that finally goes extinct, versus the time $t_0$ when this mutant appears during the environment degradation. We obtain a very good agreement between the results of our stochastic simulations and our analytical prediction in Eq. \ref{text}. For $t_0 < \theta$, mutants are less fit than wild-type organisms, and S mutants are less fit than G mutants (see Eq.~\ref{fS}). Conversely, for $t_0 > \theta$, mutants are fitter than wild-type organisms, and S mutants are fitter than G mutants: hence, S mutants are always more extreme than G mutants. Because of this, intuition based e.g. on the fixation times within the Moran process~\cite{Ewens79,Teimouri19,Teimouri19b} with constant population size make us expect that S mutants will have their fates sealed faster, and thus will get extinct faster provided that they are destined for extinction (note that related results exist in the framework of the Wright-Fisher model, see e.g.~\cite{Maruyama74}). This is indeed what we obtain (see Fig.~\ref{HE_text}). In particular, the largest extinction time is obtained close to $t_0=\theta$, where G and S mutants are neutral. In addition, for $t_0 \ll \theta$, S mutants have a fitness $f_S\approx0$ (see Eq.~\ref{fS}). Then, they generally go extinct in about one generation, i.e. in $\tau_0\approx 10$ time units (in our simulations, the death rate, which sets the division rate when the population is close to its steady-state size $K(1-g_W/f_W)$, is taken equal to 0.1): this is what is obtained in Fig. \ref{HE_text}. Still for $t_0 \ll \theta$, G mutants are such that $f_G=0.5$ while $f_W\approx1$ (see Eq.~\ref{HE_eq}): then, the extinction time of the mutant lineage can be obtained within the framework of the Moran process assuming a constant population size $K(1-g_W/f_W)$: it yields $\tau_0\approx15$~\cite{Ewens79}, consistently with Fig. \ref{HE_text}. Furthermore, Fig. \ref{HE_text}A shows that for $t_0 < \theta$, the bigger the Hill coefficient $n$ characterizing the steepness of the environment degradation (see Eq.~\ref{HE_eq}), the smaller the mean time to extinction. In particular, as long as $t_0<\theta$, we have $f_S\approx0$ and $f_W\approx1$, and therefore the results obtained just before for $t_0\ll\theta$ hold. Finally, Fig. \ref{HE_text}B shows that $\tau_0$ does not depend on the carrying capacity $K$. This can be understood from Eq.~\ref{text}, given that $p_{\mbox{\scriptsize fix}}$ is independent from $K$, as well as $\rho$, as explained in Section~\ref{probaFix}. 

\begin{figure}[h!]%
	\centering
	\includegraphics[width=\textwidth]{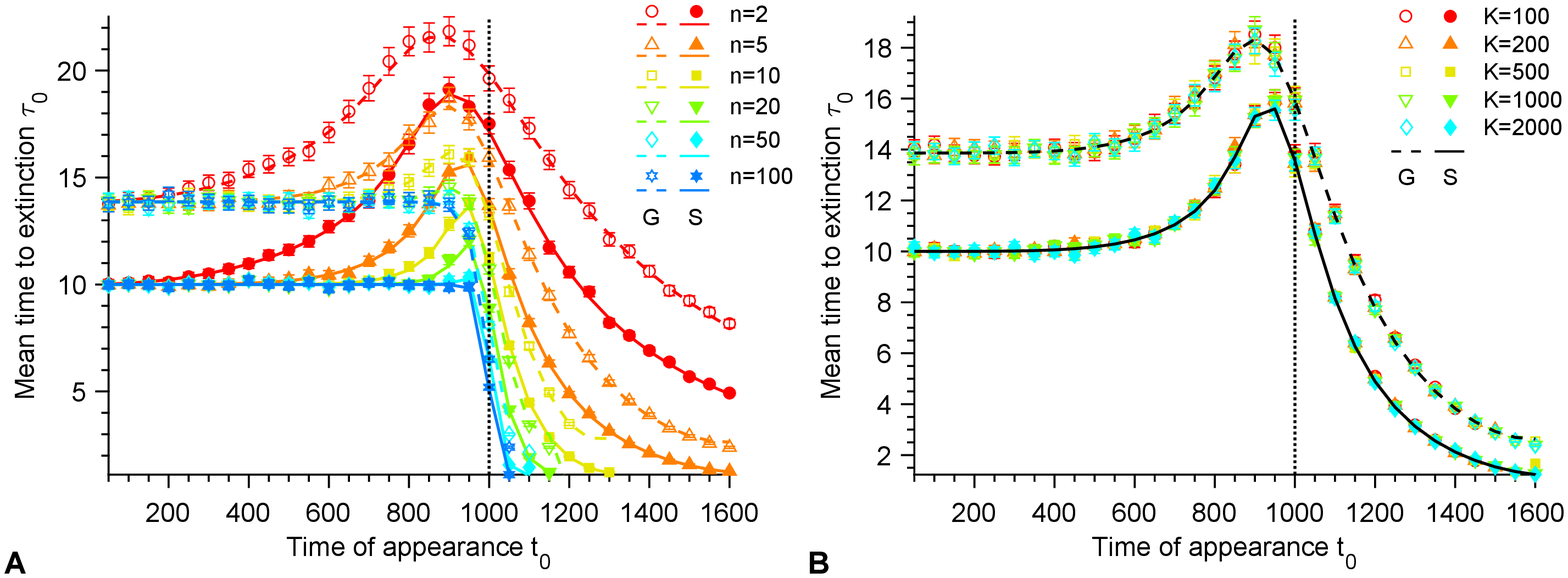}
	\vspace{0.2cm}
	\caption{{\bf Mean time to extinction.} {\bf A.} Mean time to extinction $\tau_{0}$  of G and S mutants versus their time of appearance $t_0$ in the deteriorating environment, for $K=10^3$ and for different Hill coefficients $n$ characterizing the steepness of the environment deterioration (see Eq.~\ref{HE_eq}). {\bf B.} Mean time to extinction $\tau_{0}$  of G and S mutants versus their time of appearance $t_0$ in the deteriorating environment, for different carrying capacities $K$ and a fixed Hill coefficient $n=5$ characterizing the decay of $f_W$ (see Eq.~\ref{HE_eq}). In both panels, markers correspond to averages over $10^3-10^4$ replicate stochastic simulations. Solid (resp. dashed) curves correspond to numerical integrations of Eq. \ref{text} for S (resp. G) mutants. Here, S mutants satisfy $m=n$, i.e. they have the same sensitivity to the environment as W organisms (see Eq.~\ref{fS}).  Parameter values: $g_W=g_G=g_S=0.1$, $N_W^0=10$ and $\theta=10^3$. Vertical dotted lines: $t_0=\theta$. }%
	\label{HE_text}%
\end{figure}

\newpage

\section{Analytical approximations for a sudden environment degradation}\label{Approx}

Here, we derive analytical approximations for the fixation probability $p_{\mbox{\scriptsize fix}}$, the probability $p_{\mbox{\scriptsize r}}$ of rescue and the mean time $\tau_{\mbox{\scriptsize af}}$ of appearance of a mutant that fixes in the case of a sudden environment degradation. We thus consider that the Hill coefficient $n$ describing the decay of W fitness $f_W$ tends to infinity (see Eq.~\ref{HE_eq}), as well as $m$, which describes the increase of S mutant fitness $f_S$ (see Eq.~\ref{fS}), i.e. $n,m\rightarrow \infty$. Then, the fitness transition around $t=\theta$ is very abrupt, and we therefore consider that $f_W=1$ and $f_S=0$ if $t<\theta$ while $f_W=0$ and $f_S=1$ if $t>\theta$. 

As soon as $f_W=0$, i.e. for $t>\theta$, W microbes stop dividing. In a deterministic description, their number decreases exponentially according to the function $N_W(t)=N_W^ee^{-g_W(t-\theta)}$, where $N_W^e=K(1-g_W)$ is the equilibrium size of the fully wild-type population if $f_W=1$, i.e. for $t<\theta$. For analytical convenience, we make the approximation that $N_W(t)=N_W^e$ if  $t < \theta+\tau_{1/2}$ and $N_W(t)=0$ otherwise, where $\tau_{1/2}$ is the time such that $N_W(\tau_{1/2})=K/2$ (i.e. $\tau_{1/2}=\ln(2N_W^e/K)/g_W$). While the exact choice of $\theta+\tau_{1/2}$ as a threshold is somewhat arbitrary,  it is important to choose a threshold that reflects the decay timescale of the W population. Indeed, it allows to effectively take into account the demographic pressure that mutants undergo because of the presence of W organisms during the decline of the W population. Considering a threshold $\theta$ instead of $\theta+\tau_{1/2}$ would lead one to underestimate the demographic pressure on mutants and thus to overestimate their fixation probability. Conversely, considering a threshold $\theta+\tau_0$, where $\tau_0$ is the mean time of W population extinction when W microbes no longer divide, would lead one to overestimate the demographic pressure on mutants and thus to underestimate their fixation probability. 

\subsection{Fixation probability}

\subsubsection{Generalist mutant}

Let us first focus on the fixation probability $p_{\mbox{\scriptsize fix}}^G(t_0)$ of a single generalist (G) mutant that appears at time $t_0$. Recall that the fitness of G mutants is constant. In most of our work, we take $f_G=0.5$, but here, for the sake of generality, we will retain $f_G$ in our expressions, assuming that $f_G>g_G$. Within our approximation, the fate of a mutant will strongly depend on whether $t_0<\tilde{\theta}=\theta+\tau_{1/2}$ or $t_0>\tilde{\theta}$. We start from Eq.~\ref{pfix}, which reads
\begin{equation}
p_{\mbox{\scriptsize fix}}^G( t_0)=\frac{1}{1+g_G\int_{t_0}^{\infty} e^{\rho_G(t)}\mbox{d}t}.
\label{pfixapproxG11}
\end{equation}
Two regimes need to be distinguished: 
\begin{itemize}[noitemsep,nolistsep]
\item If $t<\tilde\theta$, then $N_W(t)=K(1-g_W)$;
\item If $t \geq \tilde{\theta}$, then $N_W(t)=0$.
\end{itemize}

\phantom{hi}

For $t_0<\tilde{\theta}$, Eq.~\ref{rho} yields
\begin{equation}
\rho_G(t)=\begin{cases}
-\left(f_Gg_W-g_G\right)(t-t_0)&\mbox{ if }t_0< t<\tilde{\theta},\\
-(f_G-g_G)(t-t_0)+f_G(1-g_W)(\tilde{\theta}-t_0)&\mbox{ if }t_0 < \tilde{\theta} < t.
\end{cases}
\end{equation}
Thus, Eq. \ref{pfixapproxG11} simplifies as:
\begin{equation}
p_{\mbox{\scriptsize fix}}^G( t_0)=\frac{(f_G-g_G)(f_Gg_W-g_G)}{f_Gg_W(f_G-g_G)-e^{-(g_G-f_Gg_W)(t_0-\tilde{\theta})}f_Gg_G(1-g_W)}\mbox{ }.
\label{pfixapproxG1}
\end{equation}

For $t_0 > \tilde{\theta}$, $N_W=0$, and Eq.~\ref{rho} yields
\begin{equation}
\rho_G(t)=-\left(f_G-g_G\right)(t-t_0)\mbox{ }.
\end{equation}
Then, Eq.~\ref{pfixapproxG11} gives
\begin{equation}
p_{\mbox{\scriptsize fix}}^G(t_0)=1- g_G/f_G\mbox{ },
\label{pfixapproxG2}
\end{equation}
which corresponds to the probability that the mutant lineage survives rapid stochastic extinction in a constant-rate birth-death process, in the absence of competition~\cite{ovaskainen10, Coates18,Marrec20}. This makes sense, because within our approximation, $t_0 > \tilde{\theta}$ formally corresponds to introducing a mutant in the absence of any W individual.

Let us summarize Eqs. \ref{pfixapproxG1} and \ref{pfixapproxG2}:
\begin{equation}
p_{\mbox{\scriptsize fix}}^G(t_0)=\begin{cases}
	\frac{(f_G-g_G)(f_Gg_W-g_G)}{f_Gg_W(f_G-g_G)-e^{-(g_G-f_Gg_W)(t_0-\tilde{\theta})}f_Gg_G(1-g_W)}&\mbox{ if }t_0<\tilde{\theta}\mbox{ },\\
	1-g_G/f_G&\mbox{ if } t_0 > \tilde{\theta}\mbox{ }.
\end{cases}
\label{pfixapproxG}
\end{equation}

\subsubsection{Specialist mutant}

Let us now turn to the fixation probability $p_{\mbox{\scriptsize fix}}^S(t_0)$ of a single specialist (S) mutant that appears at time $t_0$. Again, we start from Eq.~\ref{pfix}, which reads
\begin{equation}
p_{\mbox{\scriptsize fix}}^S( t_0)=\frac{1}{1+g_S\int_{t_0}^{\infty} e^{\rho_S(t)}\mbox{d}t}.
\label{pfixapproxS11}
\end{equation}
Note that we assume $g_S<1$. Three regimes need to be distinguished: 
\begin{itemize}[noitemsep,nolistsep]
\item If $t<\theta$, then $N_W(t)=K(1-g_W)$ and $f_S(t)=0$;
\item If $\theta<t\leq\tilde{\theta}$, then $N_W(t)=K(1-g_W)$ and $f_S(t)=1$;
\item If $t \geq \tilde{\theta}$, then $N_W(t)=0$ and $f_S(t)=1$.
\end{itemize}

\phantom{hi}

If $t_0<\theta$, Eq.~\ref{rho} yields
\begin{equation}
\rho_S(t)=\begin{cases}
g_S(t-t_0)&\mbox{ if }t_0< t<\theta,\\
g_S(\theta-t_0)+(g_S-g_W)(t-\theta)&\mbox{ if }\theta < t < \tilde{\theta},\\
g_S(\theta-t_0)+(g_S-g_W)(\tilde{\theta}-\theta)+(g_S-1)(t-\tilde{\theta})&\mbox{ if } \tilde{\theta}<t\mbox{ }.
\end{cases}
\end{equation}
Note that the second term in the second and the third lines of the previous equation both vanish if $g_S=g_W$. In this case, Eq. \ref{pfixapproxS11} simplifies as:
\begin{equation}
p_{\mbox{\scriptsize fix}}^S( t_0)=\frac{e^{-g_S(\theta-t_0)}(1-g_S)}{1+g_S(1-g_S)(\tilde{\theta}-\theta)}\mbox{ }.
\label{pfixapproxS1}
\end{equation}

If $\theta < t_0 < \tilde{\theta}$, Eq.~\ref{rho} yields
\begin{equation}
\rho_S(t)=\begin{cases}
(g_S-g_W)(t-t_0) &\mbox{ if }t_0< t<\tilde{\theta},\\
(g_S-g_W)(\tilde{\theta}-t_0)+(g_S-1)(t-\tilde{\theta}) &\mbox{ if }\tilde{\theta} < t \mbox{ }.\\
\end{cases}
\end{equation}
If in addition $g_S=g_W$, Eq. \ref{pfixapproxS11} then gives
\begin{equation}
p_{\mbox{\scriptsize fix}}^S(t_0)=\frac{1-g_S}{1+g_S(1-g_S)(\tilde\theta-t_0)}\mbox{ }.
\label{pfixapproxS2}
\end{equation}

If $t_0 > \tilde{\theta}$, Eq.~\ref{rho} yields
\begin{equation}
\rho_S(t)=(g_S-1)(t-t_0)\mbox{ }.
\end{equation}
Thus, Eq. \ref{pfixapproxS11} simplifies as:
\begin{equation}
p_{\mbox{\scriptsize fix}}^S( t_0)=1-g_S\mbox{ }.
\label{pfixapproxS3}
\end{equation}
Again, this is the probability that the mutant lineage escapes rapid stochastic extinctions, in the absence of any competition. 

Let us summarize Eqs. \ref{pfixapproxS1}, \ref{pfixapproxS2} and \ref{pfixapproxS3}:
\begin{equation}
p_{\mbox{\scriptsize fix}}^S(t_0)=\begin{cases}
\frac{e^{-g_S(\theta-t_0)}(1-g_S)}{1+g_S(1-g_S)(\tilde{\theta}-\theta)}&\mbox{ if }t_0 < \theta \mbox{ },\\
\frac{1-g_S}{1+g_S(1-g_S)(\tilde\theta-t_0)}&\mbox{ if }\theta < t_0 < \tilde{\theta}\mbox{ },\\
1-g_S&\mbox{ if } \tilde{\theta} < t_0 \mbox{ }.
\end{cases}
\label{pfixapproxS}
\end{equation}

\begin{figure}[h!]
	\begin{center}
		\includegraphics[width=0.5\textwidth]{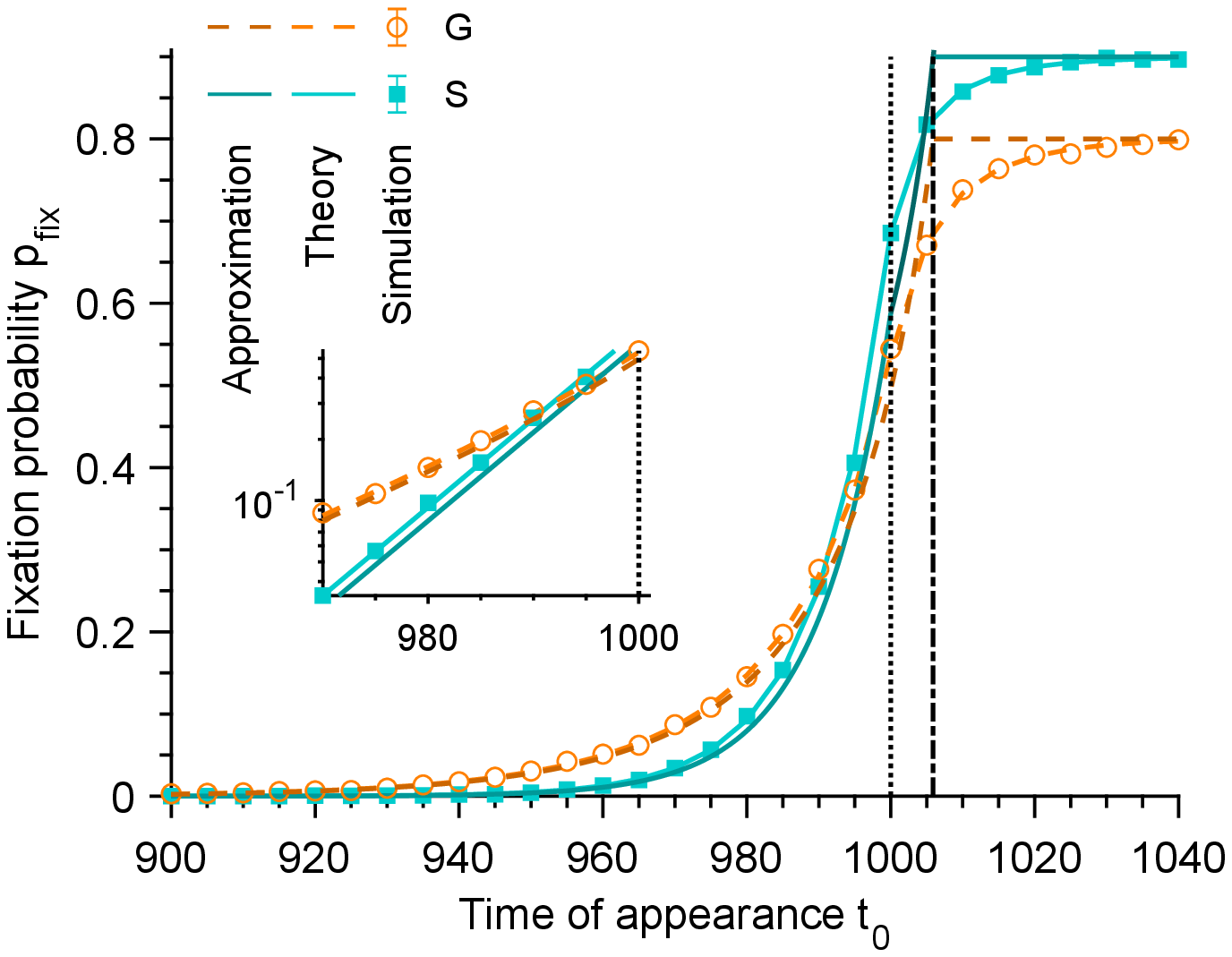}
		\caption{{\bf Fixation probability for a sudden environment degradation.} Fixation probability $p_{\mbox{\scriptsize fix}}$ of S  or G mutants versus their time of appearance $t_0$ in the deteriorating environment, for Hill coefficients $n,m\rightarrow\infty$ (see Eqs.~\ref{HE_eq} and~\ref{fS}) corresponding to an instantaneous, stepwise, environment change. 
		 Markers correspond to averages over $10^4$ replicate stochastic simulations. Light dashed (resp. solid) curves correspond to our analytical predictions in Eq. \ref{pfix} for G (resp. S) mutants. Dark dashed (resp. solid) curves correspond to our approximations in Eq. \ref{pfixapproxG} (resp. Eq.~\ref{pfixapproxS}) for G (resp. S) mutants in the different regimes discussed. Vertical dotted line: $t_0=\theta$. Vertical dash-dotted line: $t_0=\tilde{\theta}=\theta+\tau_{1/2}$. Parameter values: $g_W=g_G=g_S=0.1$, $K=10^3$, $N_W^0=10$, $n=m=10^{10}$, $\theta=10^3$ and $\tau_{1/2}=5.9$. Main panel: linear scale; inset: semi-logarithmic scale.}
		\label{Approxpfix}
	\end{center}
\end{figure}

Fig. \ref{Approxpfix} shows that Eqs.~\ref{pfixapproxG} and~\ref{pfixapproxS} provide good approximations in the appropriate regimes, i.e. for $t_0$ substantially smaller or larger than $\theta$. (Our approximation that the decay of the W population occurs instantaneously is least valid when $t_0$ is close to $\theta$.) 

\newpage

\subsection{Rescue probability}
\label{Approx_pr}

Now, let us focus on the rescue probability $p_{\mbox{\scriptsize r}}$, which satisfies $p_{\mbox{\scriptsize r}}=1-e^{-\Sigma}$ (see Eq.~\ref{paf}), where $\Sigma$ is given by Eq.~\ref{theSigma}. Since here $f_W(t)=0$ for $t > \theta$ and $f_W(t)=1$ for $t < \theta$, Eq.~\ref{theSigma} simplifies into
\begin{equation}
\Sigma=\mu N_W\left(1-\frac{N_W}{K}\right)\int_0^\theta p_{\mbox{\scriptsize fix}}(t)\mbox{d}t=\mu K(1-g_W)g_W\int_0^\theta p_{\mbox{\scriptsize fix}}(t)\mbox{d}t\mbox{ },
\end{equation}
where we have employed $N_W=K(1-g_W)$.
Thus, we obtain a simplified formula for the rescue probability:
\begin{equation}
p_{\mbox{\scriptsize r}}=1-\exp\left(-\mu K(1-g_W)g_W\int_0^\theta p_{\mbox{\scriptsize fix}}(t) \mbox{d}t\right)\mbox{ },
\label{prapprox}
\end{equation}
which holds both for generalist and for specialist mutants. 

Specifically, in the case of a generalist mutant, Eq.~\ref{pfixapproxG} yields 
\begin{equation}
\int_0^\theta p_{\mbox{\scriptsize fix}}^G(t)\mbox{d}t=\frac{1}{f_Gg_W}\log\left(\frac{g_G(1-g_W)e^{(g_G-f_Gg_W)\tilde{\theta}}-g_W(f_G-g_G)}{g_G(1-g_W)e^{(g_G-f_Gg_W)\tilde{\theta}}-g_W(f_G-g_G)e^{(g_G-f_Gg_W)\theta}}\right)\mbox{ }.
\label{intpfix}
\end{equation}

And in the case of a specialist mutant, Eq.~\ref{pfixapproxS} gives
\begin{align}
\int_0^\theta p_{\mbox{\scriptsize fix}}^S(t)\mbox{d}t=\frac{(1-e^{-g_S \theta})(1-g_S)}{g_S+g_S^2(1-g_S)(\tilde{\theta}-\theta)}\mbox{ }.
\label{intpfixS}
\end{align}

\begin{figure}[h!]
	\begin{center}
		\includegraphics[width=\textwidth]{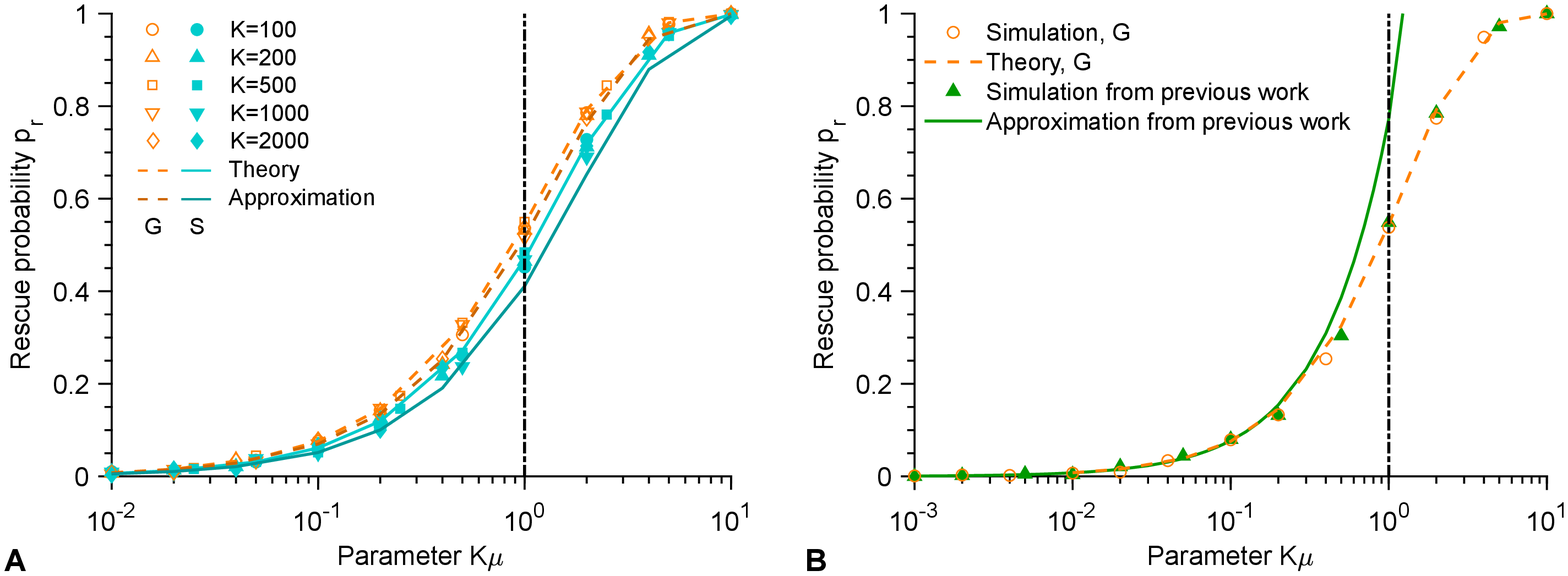}
		\vspace{0.2cm}
		\caption{{\bf Rescue probability for a sudden environment degradation.} {\bf A.} Rescue probability $p_{\mbox{\scriptsize r}}$ versus the product $K\mu$ of the carrying capacity $K$ and the mutation probability $\mu$ upon division, for different carrying capacities $K$. Markers correspond to averages over $10^4$ replicate stochastic simulations. Light dashed (resp. solid) curves correspond to our analytical predictions in Eq. \ref{paf} for G (resp. S) mutants. Dark dashed (resp. solid) curves correspond to our approximations, corresponding to Eq.~\ref{prapprox} with Eq. \ref{intpfix} (resp. Eq.~\ref{intpfixS}) for G (resp. S) mutants, with $\tau_{1/2}=5.9$. {\bf B.} Rescue probability $p_{\mbox{\scriptsize r}}$ versus $K\mu$. The present results for G mutants are compared to those of our previous work~\cite{Marrec20} for $K=10^3$. Markers correspond to averages over $10^3-10^4$ replicate stochastic simulations. Dashed orange curve: analytical prediction in Eq. \ref{paf} for G mutants. Solid green curve: analytical prediction $p_{\mbox{\scriptsize r}}=1-p_0$ with $p_0$ in Eq. 1 of~\cite{Marrec20}, valid for $K\mu\ll1$. Vertical dash-dotted lines in both panels: $K\mu=1$. Parameter values: $g_W=g_G=g_S=0.1$, $N_W^0=10$, $n=m=10^{10}$, $\theta=10^3$.}
		\label{Approxpaf}
	\end{center}
\end{figure}

Fig. \ref{Approxpaf}A shows that there is a good agreement between our approximated analytical predictions and our numerical simulation results. Moreover, we observe that the transition between small and large values of  $p_{\mbox{\scriptsize r}}$ occurs for $\mu K$ of order 1. Indeed for abrupt environment degradations such that W fitness gets to 0 right at the transition point $\theta$, preexisting mutants are necessary to ensure rescue.

In a previous work~\cite{Marrec20}, we proposed an expression for the probability of extinction of a microbial population subjected to a periodic presence of antimicrobial in the weak-mutation regime $K\mu\ll1$. We then assumed that the antimicrobial was instantaneously added and removed from the environment, which thus corresponds to instantaneous environment changes. For a perfect biostatic antimicrobial that completely stops growth, wild-type fitness goes to 0 in the presence of antimicrobial, corresponding to the case studied here. When in addition the alternation period is long enough for extinction to occur at the first phase with antimicrobial if no resistant mutants preexist, our prediction in Eq. 1 of~\cite{Marrec20} gives a good approximation of our present results, as shown by Fig.~\ref{Approxpaf}B. Therefore, the present work generalizes this prediction beyond the weak-mutation regime $K\mu\ll1$. Note that in~\cite{Marrec20} we made the assumption $K\mu\ll1$ in particular when calculating the probability that at least one mutant be present when antimicrobial is added. Indeed, we expressed it as as the ratio of the average lifetime of a mutant lineage (destined for extinction in the initial environment) to the average time of appearance of a new mutant lineage. This assumes that at most one mutant lineage is present in the population.

\newpage

\subsection{Appearance time of a mutant that fixes}

Finally, we derive an approximated analytical prediction for the mean time of appearance $\tau_{\mbox{\scriptsize af}}$ of a mutant that fixes in the population before it goes extinct. Let us recall that the probability density function of $\tilde{\tau}_{\mbox{\scriptsize af}}$ satisfies $F_{\tilde{\tau}_{\mbox{\scriptsize af}}}(t)=(1/p_{\mbox{\scriptsize r}})(\mbox{d}p_{\mbox{\scriptsize af}}/\mbox{d}t)$ (see Eq. \ref{Ftauaf} and above). Thus, for an abrupt environment degradation such that $f_W(t)=0$ for $t>\theta$, the mean time of appearance $\tau_{\mbox{\scriptsize af}}$ is given by:
\begin{equation}
\tau_{\mbox{\scriptsize af}}=\int_0^{\theta}tF_{\tilde{\tau}_{\mbox{\scriptsize af}}}(t)\mbox{d}t=\frac{1}{p_{\mbox{\scriptsize r}}}\int_0^\theta t \frac{\mbox{d}p_{\mbox{\scriptsize af}}}{\mbox{d}t}\mbox{d}t=\theta-\frac{1}{p_{\mbox{\scriptsize r}}}\int_0^\theta p_{\mbox{\scriptsize af}}(t) \mbox{d}t=\theta-\frac{1}{p_{\mbox{\scriptsize r}}}\int_0^\theta (1-e^{-\sigma(t)}) \mbox{d}t\mbox{ },
\label{tauafapprox}
\end{equation}
where we have performed an integration by parts, employed Eq.~\ref{petitsigma} (and the formula for $p_{\mbox{\scriptsize af}}(t)$ just above it), and used $p_{\mbox{\scriptsize af}}(\theta)=p_{\mbox{\scriptsize r}}$ (see Eq.~\ref{paf}, and recall that here, $f_W(t)=0$ for $t>\theta$). Using Eq.~\ref{thesigma} with $f_W=1$ and $N_W=K(1-g_W)$ for $t<\theta$, we have
\begin{equation}
\sigma(t)=\mu K g_W (1-g_W) \int_0^t p_{\mbox{\scriptsize fix}}(u)\mbox{d}u \mbox{ }.
\label{Sigmat}
\end{equation}
Eq. \ref{tauafapprox} is valid for both generalist and specialist mutants. One just needs to compute $p_r$ by using Eq. \ref{prapprox} with Eq. \ref{intpfix} (resp. Eq.~\ref{intpfixS}) for G (resp. S) mutants and $p_{\mbox{\scriptsize fix}}$ by using Eq. \ref{pfixapproxG} (resp. Eq.~\ref{pfixapproxS}) for G (resp. S) mutants.

\begin{figure}[htb]
	\centering
\includegraphics[width=0.5\textwidth]{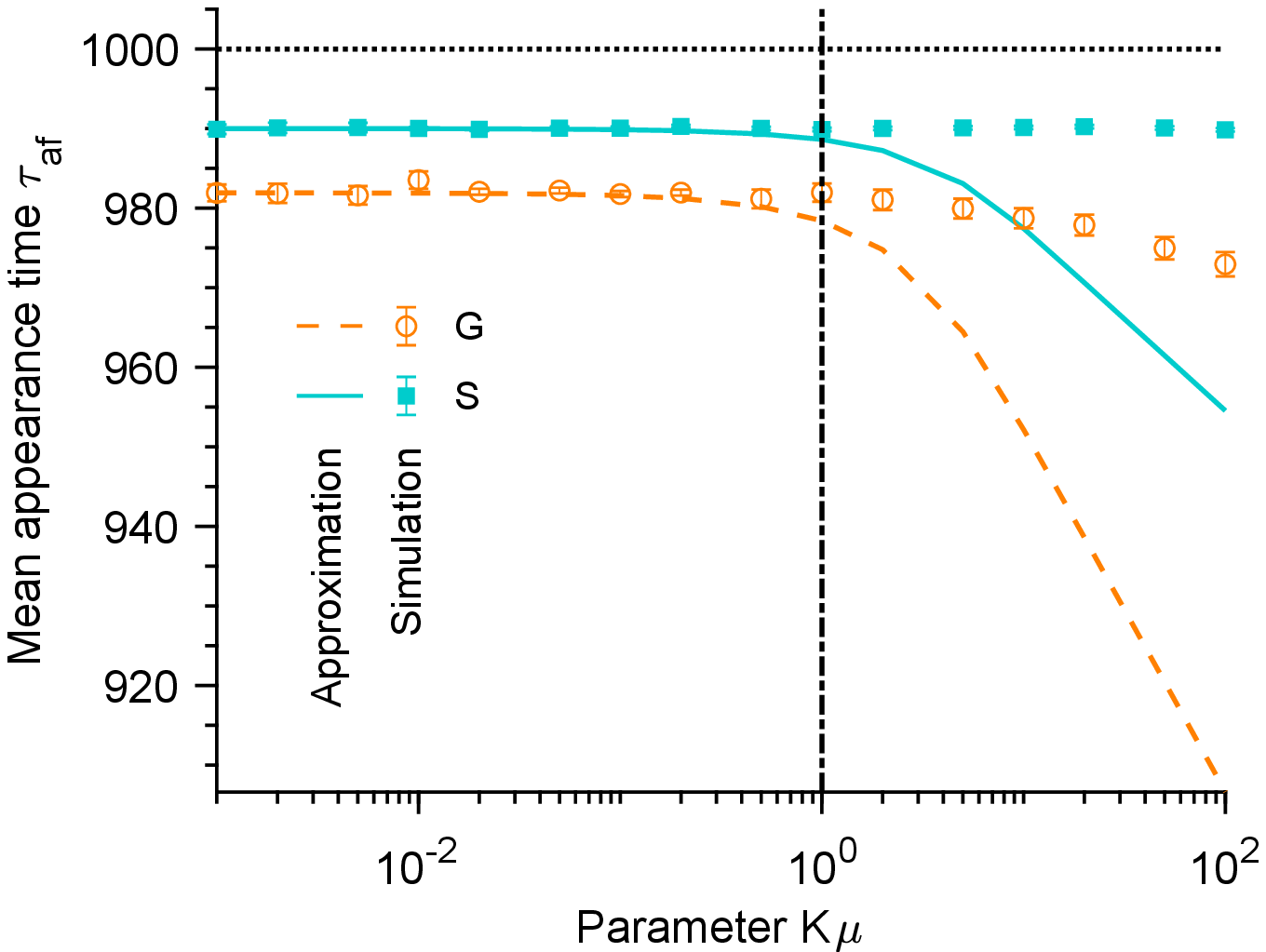}%
\vspace{0.2cm}
	\caption{{\bf Mean time of appearance for a sudden environment degradation.} Mean time $\tau_{\mbox{\scriptsize af}}$ of appearance of a G or S mutant that fixes versus the product $K\mu$ of the carrying capacity $K$ and the mutation probability $\mu$. Here, $\mu$ was varied at constant carrying capacity $K=10^3$. Horizontal dotted line: $\tau_{\mbox{\scriptsize af}}=\theta$. Vertical dash-dotted line: $K\mu=1$.
	Markers correspond to averages over $10^3$ replicate stochastic simulations (``Simulation''). Dashed and solid lines correspond to our analytical predictions (``Theory'') for G and S mutants, respectively (see Eq. \ref{tauafapprox}). Parameter values: $g_W=g_G=g_S=0.1$, $N_W^0=10$, $m=n=10^{10}$, $\theta=10^3$ and $\tau_{1/2}=5.9$ and $\theta=10^3$.
	}%
	\label{tauafapproxfig}%
\end{figure}

Fig. \ref{tauafapproxfig} shows that there is a very good agreement between our approximated analytical predictions and the results of our numerical simulations in the weak-to-moderate mutation regime $K \mu \lesssim 1$ where our analytical derivations were conducted (see main text, ``Rescue probability" section). Recall also that $\tau_{\mbox{\scriptsize af}}$ only depends on $K$ and $\mu$ via $K\mu$ (see main text).  

\newpage

\section{From the stochastic model to the deterministic limit}
\label{From the stochastic model to the deterministic limit}

In our analytical calculations, we consider the deterministic description for the population of W organisms (see Eq. \ref{dNAdt}). Here, we present a full derivation of the deterministic limit of the stochastic model for large population sizes. This derivation is similar to those of Refs.~\cite{Traulsen05,Traulsen09,Marrec18} that address the case of the Moran model.

In a fully wild-type (W) population, the probability $P(j,t|j_0)$ of having $j$ W microorganisms at time $t$, knowing that $j_0$ W microorganisms were present at time $t=0$, satisfies the master equation
\begin{align}
\frac{\partial P(j,t|j_0)}{\partial t}&=f_W(t)\left(1-\frac{j-1}{K}\right)(j-1)P(j-1,t|j_0)+g_W(j+1)P(j+1,t|j_0)\nonumber\\
&-\left[f_W(t)\left(1-\frac{j}{K}\right)+g_W\right]jP(j,t|j_0)\mbox{ }.
\end{align}
Let us introduce  $x=j/K$ and $\rho(x,t|x_0)=KP(j,t|j_0)$, and perform a Kramer-Moyal expansion \cite{VanKampen, gardiner}, which focuses on the regime $1/K\ll x$. To first order in $1/K$, one obtains the following diffusion equation~\cite{Ewens79} (also known as Fokker-Planck equation or Kolmogorov forward
equation):
\begin{equation}
\frac{\partial \rho(x,t|x_0)}{\partial t}=-\frac{\partial}{\partial x}\left\{\left[f_W(t)x(1-x)-g_Wx\right]\rho(x,t|x_0)\right\}+\frac{1}{2K}\frac{\partial^2}{\partial x^2}\left\{\left[f_W(t)x(1-x)+g_Wx\right]\rho(x,t|x_0)\right\}\mbox{ }.
\end{equation}
Note that the first term on the right hand-side of this equation corresponds to the selection term (known as the drift term in physics), while the second one corresponds to the genetic drift term (known as the diffusion term in physics). 

In the limit $K \rightarrow \infty$, to zeroth order in $1/K$, one can neglect the diffusion term, yielding:
\begin{equation}
\frac{\partial \rho(x,t|x_0)}{\partial t}=-\frac{\partial}{\partial x}\left\{\left[f_W(t)x(1-x)-g_Wx\right]\rho(x,t|x_0)\right\}\mbox{ }.
\end{equation}
In this limit, one obtains an equation on the average population size (scaled by $K$), $\langle x(t) \rangle=\int_0^1 x\rho(x,t|x_0)dx$:
\begin{equation}
\frac{\partial \langle x \rangle}{\partial t}=\left[f_W(t)-g_W\right]\langle x \rangle-f_W(t)\langle x^2 \rangle \mbox{ }.
\end{equation}
Further assuming that the distribution of $x$ is very peaked around its mean ($\langle x \rangle \approx x$) and in particular neglecting the variance ($\langle x^2 \rangle \approx \langle x \rangle^2\approx x^2$), which is acceptable for very large systems with demographic fluctuations, one obtains:
\begin{equation}
\frac{\partial x}{\partial t}=\left[f_W(t)(1-x)-g_W\right]x \mbox{ }.
\label{dxdt}
\end{equation}
Multiplying this ordinary differential equation by the carrying capacity $K$ yields Eq. \ref{dNAdt}, where $j$ is denoted by $N_W$.

\section{Numerical integration methods}
\label{Numerical resolution details}

In this work, we derived analytical predictions for the fixation probability $p_{\mbox{\scriptsize fix}}$, the rescue probability $p_r$ and the mean time of extinction $\tau'_0$ (see Eqs. \ref{pfix}, \ref{paf} and \ref{text}, respectively). Since these equations involve improper integrals, it is necessary to appropriately choose the values of the (finite) integral boundaries in order to obtain a good approximation of these improper integrals by numerical integration. These choices are discussed below. The built-in function \texttt{NIntegrate} from Wolfram Mathematica was then employed to perform numerical integrations.

\begin{figure}[h!]
	\centering
	\includegraphics[width=\textwidth]{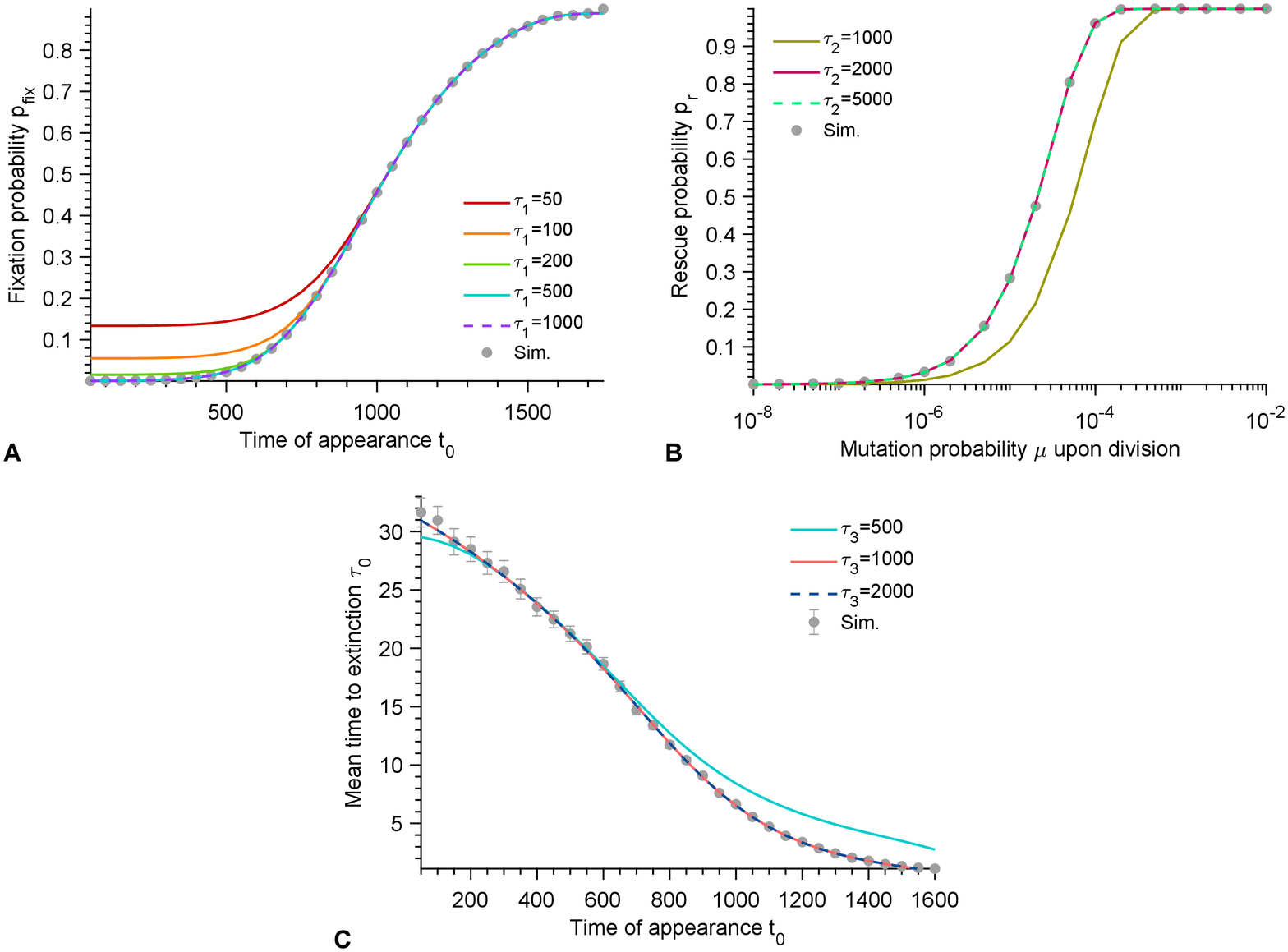}%
	\vspace{0.2cm}
	\caption{{\bf Robustness of parameters and numerical integrations.} {\bf A.} Fixation probability $p_{\mbox{\scriptsize fix}}$ of G mutants versus their time of appearance $t_0$ in the deteriorating environment. Solid curves correspond to numerical integrations of Eq. \ref{tau1} with different values of $\tau_1$. {\bf B.}  Rescue probability $p_{\mbox{\scriptsize r}}$ of a W population in a deteriorating environment by G mutants, versus mutation probability $\mu$ upon division. Solid curves correspond to numerical integrations of Eq. \ref{tau2} with different values of $\tau_2$. {\bf C.} Mean time of extinction $\tau'_{0}$ of G mutants versus their time of appearance $t_0$ in the deteriorating environment. Solid curves correspond to numerical integrations of Eq. \ref{tau3} with different values of $\tau_3$. In all panels, gray markers correspond to averages over $10^3$ replicate stochastic simulations, and error bars in panel C (often smaller than markers) to 95\% confidence intervals. Parameter values: $f_G=1$ (recall that generally we take $f_G=0.5$), $g_W=g_G=g_S=0.1$, $K=10^3$, $N_W^0=10$, $n=5$ and $\theta=10^3$. }%
	\label{Rob}%
\end{figure}

First, in order to compute numerically $p_{\mbox{\scriptsize fix}}$ from Eq. \ref{pfix}, let us introduce a parameter $\tau_1$ such that:
\begin{equation}
p_{\mbox{\scriptsize fix}}(t_0)=1-\frac{g_M\int_{t_0}^{\infty} e^{\rho(t)}\mbox{d}t}{1+g_M\int_{t_0}^{\infty} e^{\rho(t)}\mbox{d}t}\approx1-\frac{g_M\int_{t_0}^{t_0+\tau_1} e^{\rho(t)}\mbox{d}t}{1+g_M\int_{t_0}^{t_0+\tau_1} e^{\rho(t)}\mbox{d}t}\mbox{ },
\label{tau1}
\end{equation}
One should choose $\tau_1$ such that it is much larger than the mean time of extinction of the mutants $\tau'_0$. Otherwise, some mutants destined for extinction will be considered as mutants that fix. Fig. \ref{Rob}A illustrates this point: for the parameters employed in  this figure, the largest value of $\tau_0$ is $\max(\tau_0) \sim 30$, and accordingly, we observe that for $\tau_1 \gg 30$, the agreement between the analytical prediction calculated numerically via Eq.~\ref{tau1} and the simulated data is very good.

Similarly, in order to compute numerically $p_{\mbox{\scriptsize r}}$ from Eq. \ref{paf}, we introduce a parameter $\tau_2$ such that:
\begin{equation}
p_{\mbox{\scriptsize r}}=1-\exp\left[-\mu\int_0^{\infty}p_{\mbox{\scriptsize fix}}(t)N_W(t)f_W(t)\left(1-\frac{N_W(t)}{K}\right) \mbox{d}t\right]\approx1-\exp\left[-\mu\int_0^{\tau_2}p_{\mbox{\scriptsize fix}}(t)N_W(t)f_W(t)\left(1-\frac{N_W(t)}{K}\right) \mbox{d}t\right]\mbox{ },
\label{tau2}
\end{equation}
Choosing $\tau_2$ so that it is larger than the mean time of spontaneous extinction of wild-type microbes should ensure that we capture the whole time range over which mutants can appear and fix. As can be seen in Fig. \ref{HE_NA}, for the parameter values chosen in Fig. \ref{Rob}B, the mean time of spontaneous extinction is $\sim 1750$. Indeed, Fig. \ref{Rob}B shows that a good agreement between numerical predictions and simulated data is obtained for $\tau_2 > 1750$. 

Similarly, in order to compute numerically $\tau_0=\tau_0^{'}-t_0$ from Eq. \ref{text} with $i_0=1$, we introduce a parameter $\tau_3$ such that:
\begin{equation}
\tau_0^{'}=\frac{g_M}{1-p_{\mbox{\scriptsize fix}}(t_0)}\int_{t_0}^{\infty}\frac{te^{\rho(t)}}{(1+\Lambda(t))^{2}}\mbox{d}t\approx\frac{g_M}{1-p_{\mbox{\scriptsize fix}}(t_0)}\int_{t_0}^{t_0+\tau_3}\frac{te^{\rho(t)}}{(1+\Lambda(t))^{2}}\mbox{d}t \mbox{ }.
\label{tau3}
\end{equation}
The parameter $\tau_3$ must be chosen so that it is larger than all times for which the probability density function of $\widehat{\tau}_0$ is significant. In practice, we may choose $\tau_3$ as larger than the variance of the distribution of extinction times. Assuming that this distribution is exponential (it is close to exponential in simulations), one should choose $\tau_3 \gg \tau_0^2$. Accordingly, Fig. \ref{Rob}C demonstrates a very good agreement with simulated data for $\tau_3 \gg \max(\tau_0)^2 \sim 900$, where $\max(\tau_0)$ is the largest value of $\tau_0$ for the parameters involved in this figure.
 
 In practice, in each figure of this paper, we chose the values of $\tau_1$, $\tau_2$ and $\tau_3$ so that they were large enough to satisfy the criteria outlined here in the worse case of the figure (i.e. the one requiring the largest value of this parameter).
 
\clearpage

\section{Numerical simulation methods}
\label{SimuAppendix}

\begin{figure}[htb]
	\begin{center}
		\includegraphics[width=0.5\textwidth]{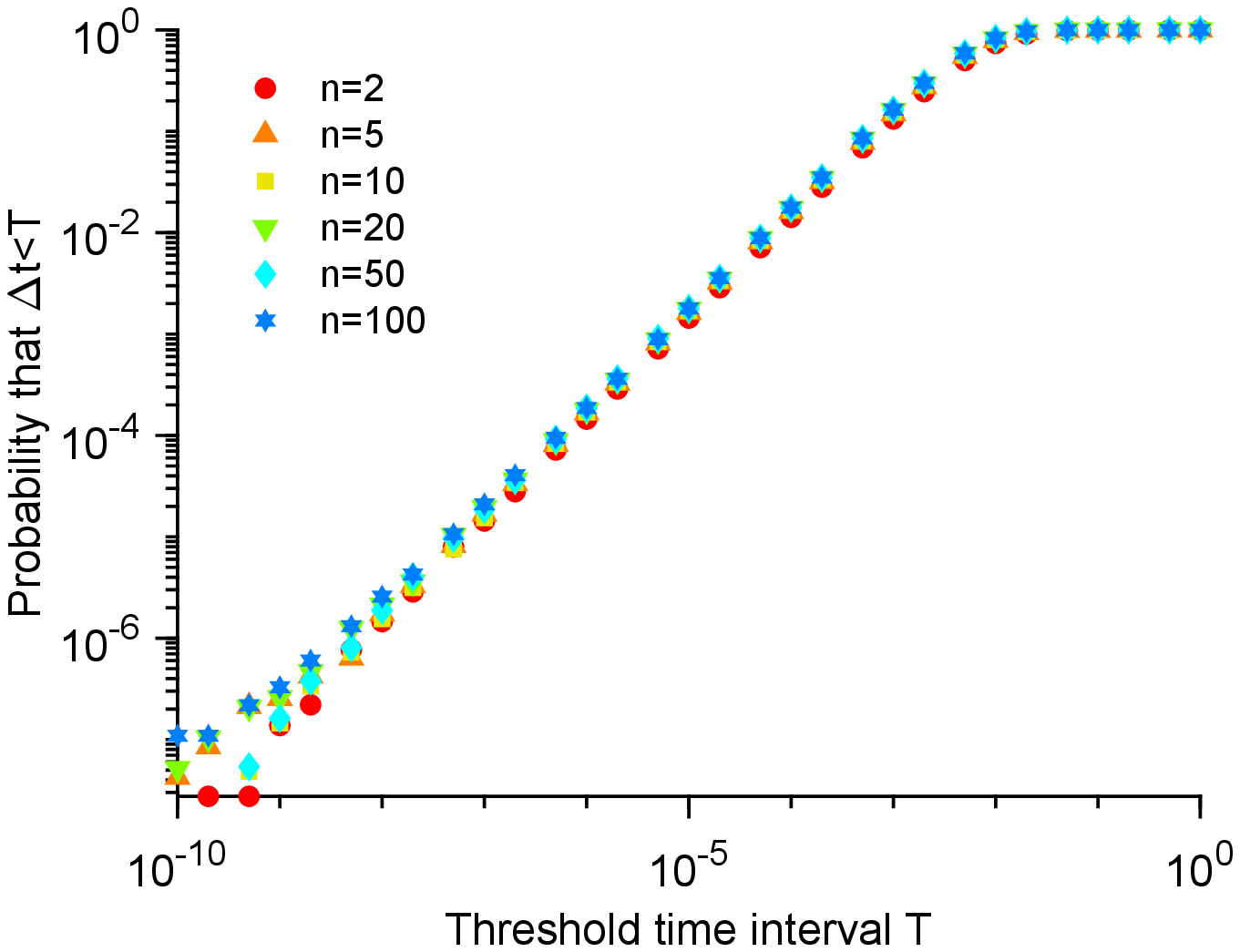}
		\vspace{0.2cm}
		\caption{{\bf Time interval between two events.} Probability that the sampled time interval $\Delta t$ between two events in the Gillespie simulation is smaller than the threshold time interval $T$ plotted versus $T$ for different Hill coefficients $n$ (see Eqs.~\ref{HE_eq}). Markers correspond to the average over $10^2$ replicate stochastic simulations of a purely $W$ population ($\mu=0$). Parameter values: $g_W=0.1$, $K=10^3$, $N_W^0=10$ and $\theta=10^3$.}
		\label{IntTime}
	\end{center}
\end{figure}

In this work, all numerical simulations are performed using a Gillespie algorithm \cite{Gillespie77}. Because the sampled time intervals $\Delta t$ between successive individual event satisfy $\Delta t<1$ (see Fig. \ref{IntTime}), which is smaller than the timescales of all processes considered here, we neglect fitness variations between individual events. In practice, the sampled time intervals between each individual event tend to get larger close to extinction events, since the total number of microbes then substantially decreases, but even then, they remain smaller than 1. Note that, in order to take into account the time variability of fitness at a higher resolution than that of events, one could employ e.g. the approach described in \cite{Thanh15}. In the following, we provide details about the simulations used in each part of our work. Matlab implementations of our numerical simulations are freely available at \url{https://doi.org/10.5281/zenodo.3993272}.

\subsection{Population decay in a deteriorating environment}
\label{NS - Hill equation}

In our simplest simulations, presented in Fig.~\ref{HE_NA}, only W microorganisms were considered (no mutation, $\mu=0$). For each replicate simulation, we saved the number of W individuals present at regular time intervals, i.e. at time points $0, \delta t, 2\delta t...$ The elementary events that can occur are:
	\begin{itemize}
		\item $W \rightarrow 2W$: Division of a wild-type microbe with rate $k_W^{+}=f_W(t)(1-N_W/K)$, where the value of $f_W(t)$ is taken at the time $t$ of the last event that occurred. 
		\item $W \rightarrow \emptyset$: Death of a wild-type microbe with rate $k_W^{-}=g_W$.
	\end{itemize}
The total rate of events is $R=(k_W^++k_W^-)N_W$. Simulation steps are the following:
\begin{enumerate}
	\item Initialization: The microbial population starts from $N_W=N_W^0$ wild-type microorganisms at time $t=0$, and the value of $f_W$ is set at $f_W(0)$. 
	\item The time increment $\Delta t$ is sampled randomly from an exponential distribution with mean $1/R$, where $R=(k_W^++k_W^-)N_W$. The next event to occur is chosen randomly, with probabilities $k/R$ proportional to the rate $k$ of each event.
	\item The time $t$ is increased to $t=t+\Delta t$ and the event chosen at Step 2 is executed, i.e. $N_W$ is updated. The value of $f_W$ is also updated from $f_W(t)$ to $f_W(t+\Delta t)$.
	\item The number of wild-type microbes $N_W$ is saved at the desired time points falling between $t$ and $t+\Delta t$.
	\item We go back to Step 2 and iterate until the total number of microbes reaches zero ($N_W=0$), corresponding to extinction.
\end{enumerate}

\subsection{Fixation probability and time of extinction of mutants}
\label{Fixation probability and time to extinction of mutants}

In our simulations concerning the fixation probability and the time of extinction of mutants, both  wild-type microorganisms (W) and mutants (M) are considered, but no random mutations are allowed, i.e. $\mu=0$. Indeed, the aim is to determine the fate of $i_0$ mutants that are introduced at a controlled time $t_0$ (generally we take $i_0=1$ to model the appearance of a single mutant). The elementary events that can occur are:
	\begin{itemize}
		\item $W \rightarrow 2W$: Division of a wild-type microbe with rate $k_W^{+}=f_W(t)(1-(N_W+N_M)/K)$, where the value of $f_W(t)$ is taken at the time $t$ of the last event that occurred.  
		\item $W \rightarrow \emptyset$: Death of a wild-type microbe with rate $k_W^{-}=g_W$.
		\item $M \rightarrow 2M$: Division of a mutant microbe with rate $k_M^{+}=f_M(t)(1-(N_W+N_M)/K)$, where the value of $f_M(t)$ is taken at the time $t$ of the last event that occurred. Note that for G mutants, $f_M$ is constant, but for S mutants, it varies in time.
		\item $M \rightarrow \emptyset$: Death of a mutant microbe with rate $k_M^{-}=g_M$.
	\end{itemize}
The total rate of events is $R=(k_W^++k_W^-)N_W+(k_M^++k_M^-)N_M$. Simulation steps are the following:
\begin{enumerate}
	\item Initialization: The microbial population starts from $N_W=N_W^0$ wild-type microorganisms and $N_M=0$ mutant at time $t=0$, and the values of $f_W$ and $f_M$ are set at $f_W(0)$ and $f_M(0)$, respectively. 
	\item The time increment $\Delta t$ is sampled randomly from an exponential distribution with mean $1/R$, where $R=(k_W^++k_W^-)N_W+(k_M^++k_M^-)N_M$. The next event to occur is chosen randomly, with probabilities $k/R$ proportional to the rate $k$ of each event.
	\item If $t+\Delta t \geq t_0$ for the first time, the time is set to $t=t_0$, $i_0$ wild-types microbes are replaced by $i_0$ mutants ($N_W=N_W-i_0$ and $N_M=N_M+i_0$) and the event determined at Step 2 is not executed. Otherwise, the time $t$ is increased to $t=t+\Delta t$ and the event determined at Step 2 is executed, i.e. $N_W$ or $N_M$ is updated. The values of $f_W$ and $f_M$ (in the case of an S mutant) are also updated.
	\item We go back to Step 2 and iterate until the total number of microbes is zero ($N_W+N_M=0$), corresponding to extinction of the population, or there are only mutants ($N_W=0$ and $N_M \neq 0$). In the latter case, we also check that the mutant lineage does not undergo rapid stochastic extinction by assessing whether it dies out or not before reaching a size of 100 individuals. If it reaches such a size, we consider that fixation of the mutant has occurred.
\end{enumerate}

\subsection{Rescue of a population by mutants}
\label{Appearance and fixation probability of mutants}

Finally, our simulations concerning the rescue of a population by mutants, both  wild-type microorganisms (W) and mutants (M) are considered, with a probability $\mu$ of mutation from W to M upon division. The elementary events that can occur are:
	\begin{itemize}
		\item $W \rightarrow 2W$: Division without mutation of a wild-type microbe with rate $k_W^{+}=f_W(t)(1-(N_W+N_M)/K)(1-\mu)$, where the value of $f_W(t)$ is taken at the time $t$ of the last event that occurred.   
		\item $W \rightarrow W+M$: Division with mutation of a wild-type microbe with rate $k_{WM}=f_W(t)(1-(N_W+N_M)/K)\mu$.
		\item $W \rightarrow \emptyset$: Death of a wild-type microbe with rate $k_W^{-}=g_W$.
		\item $M \rightarrow 2M$: Division of a mutant microbe with rate $k_M^{+}=f_M(t)(1-(N_W+N_M)/K)$, where the value of $f_M(t)$ is taken at the time $t$ of the last event that occurred. Note that for G mutants, $f_M$ is constant, but for S mutants, it varies in time.
		\item $M \rightarrow \emptyset$: Death of a mutant microbe with rate $k_M^{-}=g_M$.
	\end{itemize}
The total rate of events is $R=(k_W^++k_W^-+k_{WM})N_W+(k_M^++k_M^-)N_M$. Simulation steps are the following:
\begin{enumerate}
	\item Initialization: The microbial population starts from $N_W=N_W^0$ wild-type microorganisms and $N_M=0$ mutant at time $t=0$, and the values of $f_W$ and $f_M$ are set at $f_W(0)$ and $f_M(0)$, respectively. 
	\item The time increment $\Delta t$ is sampled randomly from an exponential distribution with mean $1/R$, where $R=(k_W^++k_W^-+k_{WM})N_W+(k_M^++k_M^-)N_M$. The next event to occur is chosen randomly, with probabilities $k/R$ proportional to the rate $k$ of each event.
	\item The time $t$ is increased to $t=t+\Delta t$ and the event determined at Step 2 is executed, i.e. $N_W$ and $N_M$ are updated. The value of $f_W$ and $f_M$ (in the case of an S mutant) are also updated.
	\item We go back to Step 2 and iterate until the total number of microbes is zero ($N_W+N_M=0$), corresponding to extinction of the population, or there are only mutants ($N_W=0$ and $N_M \neq 0$), corresponding to fixation of the mutant and rescue of the population.
\end{enumerate}

\end{document}